\documentclass[11pt]{article}
\usepackage{amssymb,amsmath,amsfonts,amsthm, amscd, mathrsfs,helvet}
\usepackage{bm}
\usepackage{graphicx,verbatim}
\usepackage{psfrag}
\usepackage[all]{xy}

\def\Xint#1{\mathchoice
   {\XXint\displaystyle\textstyle{#1}}%
   {\XXint\textstyle\scriptstyle{#1}}%
   {\XXint\scriptstyle\scriptscriptstyle{#1}}%
   {\XXint\scriptscriptstyle\scriptscriptstyle{#1}}%
   \!\int}
\def\XXint#1#2#3{{\setbox0=\hbox{$#1{#2#3}{\int}$}
     \vcenter{\hbox{$#2#3$}}\kern-.5\wd0}}

\def\dashint{\Xint-}
\def\Xint#1{\mathchoice
   {\XXint\displaystyle\textstyle{#1}}%
   {\XXint\textstyle\scriptstyle{#1}}%
   {\XXint\scriptstyle\scriptscriptstyle{#1}}%
   {\XXint\scriptscriptstyle\scriptscriptstyle{#1}}%
   \!\int}
\def\XXint#1#2#3{{\setbox0=\hbox{$#1{#2#3}{\int}$}
     \vcenter{\hbox{$#2#3$}}\kern-.5\wd0}}

\def\dashint{\Xint-}

\def\cO{{\mathcal{O}}}
\def\mbR{{\mathbb{R}}}

\def\cL{{\mathcal{L}}}

\def\cI{{\mathcal{I}}}

\def\pa{\partial}

\def\r2{{\sqrt{2}}}

\def\Xp{X^{+}}
\def\Xm{X^{-}}
\def\Yp{Y^{+}}
\def\Ym{Y^{-}}
\def\xp{x^{+}}
\def\xm{x^{-}}
\def\yp{y^{+}}
\def\ym{y^{-}}
\def\fsu{{\mathfrak{su}}}

\def\fsp{{\mathfrak{sp}}}
\def\fpsu{{\mathfrak{psu}}}
\def\p{\phi}
\def\tp{\tilde{\phi}}
\def\del{\delta}
\def\cX{{\mathcal{X}}}
\def\iS{{\mathit{S}}}

\def\sx{\sqrt{x}}
\def\sy{\sqrt{y}}
\def\Li2{{\rm{Li}_{2}}}
\def\tpsi{\tilde{\psi}}
\def\zp{z^{+}}
\def\zm{z^{-}}
\def\nn{\nonumber}
\def\bX{{\bf X}}

\def\bZ{{\bf Z}}
\def\tF{\tilde{F}}
\def\g{{\bf g}}
\def\SU{{\rm SU}}
\def\Sp{{\rm Sp}}

\def\bbR{{\mathbb{R}}}
\def\is{{\mathit{s}}}
\def\his{\hat{\mathit{s}}}
\def\vp{\varphi}

\begin{document}
\begin{titlepage}
\begin{flushright}
{\bf \today} \\
		DAMTP-07-51\\
		arXiv:0707.0668 [hep-th] \\
		\end{flushright}
		\begin{centering}
		\vspace{.2in}

{\large {\bf Quantum Scattering of Giant Magnons }}

\vspace{.3in}

Heng-Yu Chen, Nick Dorey and Rui F. Lima Matos\\
		\vspace{.2 in}
		DAMTP, Centre for Mathematical Sciences \\
			University of Cambridge, Wilberforce Road \\
			Cambridge CB3 0WA, UK \\

			\vspace{.4in}

{\bf Abstract} \\

We perform a first-principles semi-classical computation of the
one-loop corrections to the dispersion relation and S-matrix of Giant
Magnons in $AdS_{5}\times S^{5}$ string theory. The results agree
exactly with expectations based on the strong coupling expansion of 
the exact Asymptotic Bethe Ansatz equations. In particular we
reproduce the Hernandez-Lopez term in the dressing phase.    
\end{centering}

%\vspace{.05in}
%\baselineskip=.3in
\end{titlepage}

%%%%%%%%%%%
\section{Introduction}\label{Intro}
\paragraph{}
Recent developments in the study of planar ${\cal N}=4$ SUSY
Yang-Mills (and the dual string theory on $AdS_{5}\times S^{5}$) have
culminated in a proposal for a set of Asymptotic Bethe Ansatz
Equations (ABAE) \cite{Smatrix2004, BeiStau, 
BeisertSU22,BeisertLong,BES}. These
equations determine the exact scaling
dimensions $\Delta$, of all operators in a limit where a conserved
R-charge $J$ becomes infinite, with the difference $\Delta-J$ and the
't Hooft coupling $\lambda=g^{2}_{YM}N$ held fixed. The proposed
equations hold for all values of $\lambda$, but for $\lambda>>1$ their
predictions can be compared to the results of semiclassical
calculations in the worldsheet theory of the $AdS_{5}\times S^{5}$
string. In this limit, the basic excitations of the worldsheet theory
are solitons known as ``Giant Magnons'' which propagate on an infinitely
long string \cite{HM}. The exact ABAE lead to non-trivial predictions for the
dispersion relation of these solitons and also for their scattering
matrix. These predictions were compared to the results of a
leading-order semiclassical calculation in \cite{HM} (see also
\cite{CDO2,Roiban}). 
\paragraph{}
The main aim of this paper is to extend this comparison by performing
a first-principles calculation of the soliton dispersion relation and
S-matrix \cite{Smatrix2004} 
to the next order\footnote{In the following, we will refer to
the first two orders in the semiclassical expansion as tree-level and
one-loop respectively.} in the semiclassical expansion of the
worldsheet theory. Our main result is a complete calculation of the
soliton S-matrix at one-loop, which yields exact agreement with the
predictions of the ABAE. In particular, we will reproduce in full the
Hernandez-Lopez (HL) term in the magnon S-matrix which was originally
obtained by considering the one-loop quantum correction to a circular
string in $AdS_{5}\times S^{5}$ \cite{HL,FK2}. 
Our calculation, therefore provides
further confirmation of the {\em universality} of the HL term in
semiclassical string physics on $AdS_{5}\times S^{5}$. For other
interesting recent work on one-loop corrections, including a
derivation of the HL term in the context of finite gap solutions see
\cite{GV1,GV2}\footnote{We comment further on the relation of our
  calculation to the approach of these references at the end of this
  Section.} (see also \cite{FT} and \cite{oneloop}). 
In the rest of this introductory section we will
review some basic features of semiclassical soliton quantisation
\cite{Dashen,FK,Raj} required for our analysis.   
\paragraph{}
For simplicity we begin by considering the theory of a single scalar
field $\varphi(x,t)$ of mass $m$ in one space and one time dimension
with a dimensionless coupling constant $1/g$. The field obeys
non-linear equations with a two-parameter family of soliton solutions,
\begin{equation}
	\varphi=\varphi_{cl}(x,t;x^{(0)},p)
	\label{phicl}
\end{equation}
The soliton is a localised lump of energy density $\mathcal{E}(x,t)$ 
centred around the
point $x=x^{(0)}$ at time $t=0$ (see Figure 1).  The parameter $p$
corresponds to the conserved momentum conjugate to the spatial
coordinate $x$. The soliton has finite classical energy
$E(p)=gE_{cl}(p)$ and moves at constant velocity $v=v(p)\sim dE/dp$.
At time $t$, the energy density is therefore centred around the point
$x=x^{(0)}+vt$. All these features are realised, for example, in the
specific case of the sine-Gordon kink.  
\begin{figure}
\centering
\psfrag{A}{\footnotesize{$\mathcal{E}$}}
\psfrag{B}{\footnotesize{$x$}}
\psfrag{X}{\footnotesize{$x^{(0)}$}}
\psfrag{P}{\footnotesize{$p$}}
\includegraphics[width=100mm]{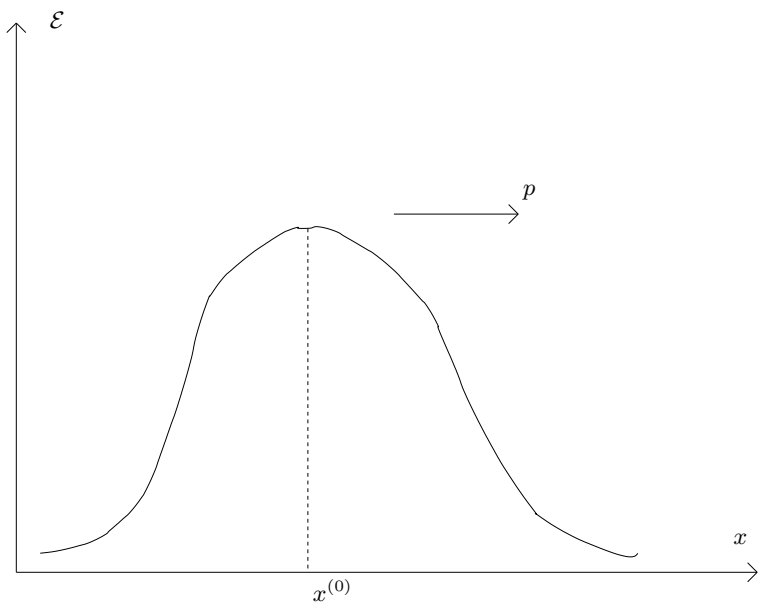}
\caption{The one-soliton solution at time $t=0$.}
\label{HLfig1}
\end{figure}
To match as closely as possible the case of interest, we will
{\it not} assume  $(1+1)$-dimensional Lorentz invariance for the full
non-linear equations of motion \footnote{In the case of the string
world-sheet theory in static gauge, two dimensional Lorentz invariance
is broken by the Virasoro constraints.}. Thus the solution
$\varphi_{cl}(x,t;x^{(0)},p)$ is not related in a simple way to the solution
with $p=0$. However, again motivated by the specific problem of
interest, we will assume that the the {\em linearised} equation of
motion takes the standard relativistic form
$(-\partial^{2}_{t}+\partial^{2}_x + m^{2})\varphi(x,t)=0$. It follows
that the soliton configuration has exponentially decaying asymptotics 
at left and right spatial infinity, 
\begin{eqnarray}
	\varphi_{cl}(x,t;x^{(0)},p)\sim \,\, \exp(-c|x|) & \qquad{} & \, {\rm
	as}\qquad{} x\rightarrow \pm \infty\,,
\end{eqnarray}
where $c=c(p)$ is a positive constant which is equal to the mass $m$
for a static soliton at rest.  
\paragraph{}
After quantisation, the soliton yields a massive single-particle
asymptotic state of the theory. Its dispersion relation has a
semiclassical expansion of the form, 
\begin{equation}
	E(p)=gE_{cl}(p)\,\,+\,\,\Delta E(p)\,\,+\,\,O\left(\frac{1}{g}\right)\,.
\end{equation} 
Our first goal is to calculate the one-loop correction to the energy
$\Delta E(p)$. In general, one-loop quantum corrections are determined
by the spectrum of the small fluctuation operator, 
\begin{equation}
	\hat{H}= \left. \frac{\delta^{2} {\cal
	L}}{\delta\varphi(x,t)^{2}}\right|_{\varphi=
	\varphi_{cl}(x,t;x^{(0)},p)}
\end{equation}
where $\mathcal{L}$ is the Lagrangian density of the theory. 
In particular we will study the auxiliary Schrodinger problem defined
by the linearised equation of motion in the soliton background, 
\begin{equation}
	\hat{H}\psi(x,t)=0
	\label{lem}
\end{equation}
where we will consider complex solutions $\psi\in \mathbb{C}$.
\paragraph{}
The asymptotics of $\hat{H}$ are determined by the asymptotics of the
soliton solution to be, 
\begin{eqnarray}
	\hat{H} & \rightarrow & -\partial^{2}_{t}+\partial^{2}_x + 
	m^{2} \,\,\, +\,\,\,O(e^{-c|x|})
\end{eqnarray}
for $x\rightarrow \pm \infty$ at fixed time $t$. For each $k\in
\mathbb{R}$, we can choose a solution, $\psi_{k}(x,t)$ of the
small fluctuation equation  (\ref{lem}) which goes like, 
\begin{eqnarray}
	\psi(x,t)& \sim & \exp\left(iE(k)t+ikx\right)
	\label{asympL}
\end{eqnarray}
with $E(k)=\sqrt{k^{2}+m^{2}}$, as $x\rightarrow -\infty$. This
corresponds to a plane-wave with wave number $k$ incident upon the
soliton from the left. Following the same solution to the 
asymptotic region to the right of the soliton, we will find that 
the solution will consist of a
transmitted wave of the form,  
\begin{eqnarray}
	\psi(x,t)& \sim & \exp \left(i\delta(k;p)\right)
	\exp\left(iE(k)t+ikx\right)
	\label{asympR}
	\end{eqnarray}
as $x\rightarrow +\infty$, where the real quantity $\delta(k;p)$
corresponds to the phase shift due to scattering on the soliton
background. Of course in a general scattering problem, to obtain
asymptotics of the form (\ref{asympR}) we would also have to include a
reflected wave which modifies the left asymptotics (\ref{asympL}). A
special feature of many integrable partial differential equations with
soliton solutions and, in particular, of the cases considered in this
paper, is that the classical reflection amplitude vanishes. Another
potential complication is the existence of normalisable bound state
solutions of the linearised equation (\ref{lem}) with exponentially
decaying asymptotics.  Again, this feature will be absent in all the
cases considered in this paper.  
\paragraph{}
The quantity $\delta(k;p)$ describes the {\em classical} scattering 
of a plane wave off the soliton background. As we now review, this 
classical scattering data is the basic ingredient we need to compute
one-loop {\em quantum} corrections to the soliton. In particular, the phase
shift $\delta(k;p)$ determines the density of scattering states which
provides the measure for integrating over the continuous spectrum of
the small fluctuation operator $\hat{H}$. The resulting formula for
the one-loop correction to the soliton energy is \cite{Dashen},
\begin{equation}
	\Delta E(p)=\frac{1}{2\pi}\, \int_{-\infty}^{+\infty} \, dk
	\frac{\partial \delta(k;p)}{\partial k}\, \sqrt{k^{2}+m^{2}}\,.
	\label{f1}
\end{equation}
The derivation of this formula is given in Appendix A. 
\paragraph{}
In the following we will need a slight generalisation to the case of $N_{F}$
scalar fields $\varphi_{I}$, $I=1,2,\ldots,N_{F}$, with Bose/Fermi
statistics depending on the sign $(-1)^{F_{I}}$. We will assume that
fluctuations of each these fields around the soliton background have
the same asymptotic dispersion relation $E=\sqrt{k^{2}+m^{2}}$ and
that the classical scattering matrix of the fluctuations is diagonal with
eigenvalues $\exp(i\delta_{I}(k;p))$, $I=1,2,\ldots ,N_{F}$. All these
features will be present in the case of interest below. With these
assumptions, the one-loop correction to the dispersion relation
becomes, 
\begin{equation}
	\Delta E(p)=\frac{1}{2\pi}\,\sum_{I=1}^{N_{F}} \, (-1)^{F_{I}} 
	\int_{-\infty}^{+\infty} \, dk
	\frac{\partial\delta_{I}(k;p)}{\partial k}\, \sqrt{k^{2}+m^{2}}\,.
	\label{f2}
\end{equation}    
In general the formulae (\ref{f1},\ref{f2}) may suffer from UV
divergences which require regularisation. In the supersymmetric case
of interest, we will find that these divergences cancel between Bosons
and Fermions.  
\paragraph{}
A characteristic feature of integrable PDEs in two spacetime
dimensions is the existence of exact classical solutions describing
the scattering of an arbitrary number of solitons. Here we will focus
on a solution describing the scattering of two solitons of momenta
$p_{1}$ and $p_{2}$ (see Figure 2),  
\begin{equation}
	\varphi=\varphi_{scat}(x,t;x^{(0)}_{1},x^{(0)}_{2},p_{1},p_{2})\,.
	\label{phiscat}
\end{equation}
\begin{figure}
\centering
\psfrag{A}{\footnotesize{$\mathcal{E}$}}
\psfrag{B}{\footnotesize{$x$}}
\psfrag{X1}{\footnotesize{$x^{(0)}_{1}$}}
\psfrag{P1}{\footnotesize{$p_{1}$}}
\psfrag{X2}{\footnotesize{$x^{(0)}_{2}$}}
\psfrag{P2}{\footnotesize{$p_{2}$}}
\includegraphics[width=100mm]{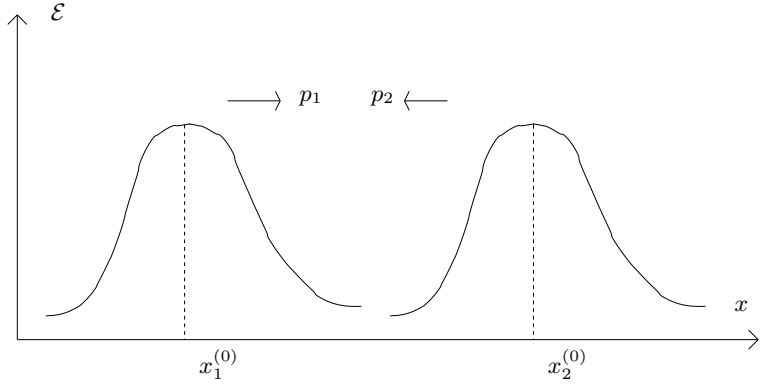}
\caption{The two-soliton scattering solution at time $t=0$.}
\label{HLfig2}
\end{figure}
As shown, the solution also depends on the positions $x^{(0)}_{1}$ and
$x^{(0)}_{2}$ of the two constituent solitons at time $t=0$. The conservation
of the higher conserved charges implied by integrability ensures that
the only effect of the scattering is a time delay $\Delta
T(p_{1},p_{2})$ relative to free propagation of the two constituent
solitons. Thus, in the far past $t\rightarrow -\infty$, and the far
future, $t\rightarrow +\infty$ the solution asymptotes to a linear
superposition of two single soliton solutions, 
\begin{equation}
	\varphi_{scat}(x,t;x^{(0)}_{1},x^{(0)}_{2},p_{1},p_{2})\rightarrow 
	\varphi_{cl}(x,t;x^{\pm}_{1},p_{1})+
	\varphi_{cl}(x,t; x^{\pm}_{2},p_{2})
\end{equation}
where the asymptotic values of the position parameters\footnote{The
parameters $x_{1}^{\pm}$ and $x_{2}^{\pm}$ as defined in this equation
should not be confused with the spectral parameters introduced later
in the paper.} as $t \rightarrow \pm \infty$ are, 
\begin{eqnarray}
	x_{1}^{\pm}=x^{(0)}_{1}\mp v_{1}\frac{\Delta T}{2}\,, & \qquad{} & 
	x^{\pm}_{2}=x^{(0)}_{2}\mp v_{2}\frac{\Delta T}{2} 
\end{eqnarray}
and, as above, the individual soliton velocities are $v_{i}\sim
dE_{i}/dp_{i}$ for $i=1,2$.  
\paragraph{} 
Another important
consequence of integrability is the factorisation of the scattering
data corresponding to the two soliton solution. In particular, a
plane-wave of wave number $k$ incident on the two-soliton solution
from the left experiences a phase shift, 
\begin{equation}
	\delta(k;p_{1},p_{2})= \delta(k;p_{1})+\delta(k;p_{2})\,.
\end{equation}
In other words, the phase shift experienced by the incident wave is
simply the sum of the phase-shifts associated with the two constituent
solitons. This property, which we will verify directly below, is
related to the complete factorisation of the S-matrix which is a
hallmark of an integrable theory. 
\paragraph{}
In the quantum theory, solitons correspond to asymptotic states which
scatter with an S-matrix, 
\begin{equation}
	{\cal S}(p_{1},p_{2})=\exp\left(i\Theta(p_{1},p_{2})\right)\,.
\end{equation}
At weak coupling, $1/g\ll1$, the phase 
$\Theta$ has a semiclassical expansion of the form, 
\begin{equation}
	\Theta(p_{1},p_{2})=g\Theta_{cl}(p_{1},p_{2})\,\,+\,\,\Delta 
	\Theta(p_{1},p_{2})\,\,+\,\,
	O\left(\frac{1}{g}\right)\,.\label{expTheta}
\end{equation} 
A famous formula of Jackiw and Woo \cite{JW} relates the leading
semiclassical contribution to the quantum S-matrix and the time-delay
$\Delta T(p_{1},p_{2})$ in classical scattering, 
\begin{equation}
	\Theta_{cl}(p_{1},p_{2})= 
	\frac{1}{g}\int_{E_{\rm Th}}^{E(p_{1},p_{2})}\Delta T(E) dE
\end{equation}
where $E_{\rm Th}$ denotes the threshold energy for scattering. Much
less well known, is the simple formula which determines the one-loop
correction to the S-matrix in an integrable field theory in terms of
the classical scattering data, 
\begin{equation}
	\Delta \Theta(p_{1},p_{2})=\frac{1}{2\pi}\,
	\int_{-\infty}^{+\infty} \, dk
	\frac{\partial \delta(k;p_{1})}{\partial k}\,\delta(k;p_{2})\,.
	\label{f3}
\end{equation}
This formula was first obtained in the context of sine-Gordon theory
by Faddeev and Korepin \cite{FK}\footnote{See, in particular, Eqn
(4.6) on p62 of this reference and the discussion following Eqn (4.28)
on p66.}. A more general derivation is provided in Appendix B.  Again
we will also require a generalisation to the case of $N_{F}$ fields
with diagonal scattering (see equation (\ref{f2})),  
\begin{equation}
	\Delta \Theta(p_{1},p_{2})=\frac{1}{2\pi}\,
	\sum_{I=1}^{N_{F}} \, (-1)^{F_{I}} 
	\int_{-\infty}^{+\infty} \, dk
	\frac{\partial\delta_{I}(k;p_{1})}{\partial k}\,\delta_{I}(k;p_{2})\,.
	\label{f4}
\end{equation}
\paragraph{}
The above formulae (\ref{f2}) and (\ref{f4}) reduces the problem of
computing one-loop corrections to the soliton dispersion and S-matrix
to the problem of finding the {\em classical} phase shifts,
$\delta_{I}(k;p)$, of small fluctuations around the background of a
{\em single} soliton. The bulk of the paper is devoted to solving this
problem for the case of a Giant Magnon solution of the worldsheet
theory of arbitrary charge. In fact, we will describe three different
approaches to determining the phase shifts. The first method is 
originally due to a clever observation of Dashen, Hasslacher and
Neveu \cite{Dashen}, that a linearised fluctuation around a background
containing $n$ solitons can be obtained as a degenerate limit of an
$n+1$ soliton solution. In particular we will apply this approach to
the exact multi-soliton solutions of the bosonic world sheet fields
constructed via the dressing method developed in \cite{Dressing,
  Dressing1}. 
The second approach relies on obtaining the spectral data for fluctuations
around the Giant Magnon in the finite-gap formalism of
\cite{KMMZ,BeiSak}. This approach reproduces the results of the 
dressing method for the
bosonic fluctuations and also produces explicit results for the phase
shifts of the fermionic fields. Finally, we provide a further check on
the phase shifts by comparing them with the proposed exact magnon
S-matrix \cite{BeisertSU22} 
in a limit where one magnon becomes a worldsheet soliton and
the other becomes an elementary fluctuation of the worldsheet fields.
Note that this comparison involves only the leading-order piece of the
proposed exact S-matrix in the semiclassical limit which already has
many independent tests. Having extracted the classical scattering
data, we use it to calculate the one-loop correction to the S-matrix
of two giant magnons using formula (\ref{f4}) and compare with the
Hernandez-Lopez one-loop contribution \cite{HL} to the exact
S-matrix. We also demonstrate the vanishing of the one-loop correction
to the soliton energy, completing an earlier partial calculation appearing in 
\cite{Spradlin}.
\paragraph{}
The paper is organised as follows. In the next Section we review the
predictions for soliton scattering coming from the ABAE. In Section 3 
we describe the different approaches to extracting the classical
scattering data outlined above. Finally in Section 4 we complete the
calculation of the one-loop corrections to the soliton dispersion
relation and S-matrix obtaining exact agreement with the predictions
described in Section 2. Various technical details and derivations are
relegated to the Appendices.      
\paragraph{}
In interesting recent work, 
Gromov and Vieira \cite{GV1,GV2} have also provided a
semiclassical derivation of the Hernandez-Lopez phase. Our calculation 
differs from theirs in that we are directly computing the S-matrix
for soliton scattering with vacuum boundary conditions, 
while they are computing the one-loop energy shift for finite gap
solutions with periodic boundary conditions. Nevertheless it is clear
that the two calculations are related. In particular, in 
Section 3.2, we obtain the classical scattering data for the
fermionic worldsheet fields in the soliton background 
by taking a limit of an appropriate finite
gap solution. On the other hand, the scattering data for the bosonic
worldsheet fields is obtained in Section 3.1 
by explicit construction of soliton scattering solutions.   
 
\section{Predictions from Bethe Ansatz and Scattering
Matrix}\label{Prediction}

\subsection{The asymptotic spectrum and its semiclassical
limits}\label{Limits}
\paragraph{}
The asymptotic spectrum of the gauge theory spin chain consists of an
infinite tower of BPS states labelled by a positive integer $Q\,,~Q\in
{\mathbb{Z}}^{+}$, and their conserved momentum $p$. The elementary
excitation,  called the ``magnon'', corresponds to the case $Q=1$.
States with $Q>1$  correspond to the bound states of these elementary
magnons \cite{Dorey}.  Being short representations of the extended
residual symmetry algebra $\fpsu(2|2)^{2}\ltimes\bbR^{3}$ which carry
conserved central charges, the dispersion relation of the elementary
magnon and the bound states is then fixed by the shortening condition
to be  \cite{ BeisertSU22, BeisertLong, Dorey, BDS},
\begin{equation}
	\Delta-J=E=\sqrt{Q^{2}+16g^{2}\sin^{2}\left(\frac{p}{2}\right)}\,. 
	\label{MagDisp}
\end{equation}
Here we have introduced the coupling
$g^{2}=\lambda/16\pi^{2}$. The magnon dispersion relation
(\ref{MagDisp}) is common to all states in the supermultiplet of
dimension  $16Q^{2}$ \cite{CDO3}.  
\paragraph{}
As usual we introduce a convenient representation of the dispersion
relation in terms of spectral parameters $X^{\pm}$ where,
\begin{equation}
	p(X^{\pm})=\frac{1}{i}\log\left(\frac{\Xp}{\Xm}\right)\,;\label{Defp}
\end{equation}
so that the energy $E$ and charge $Q$ can be expressed as
\begin{eqnarray}
	E(X^{\pm})&=&\frac{g}{i}\left[\left(\Xp-\frac{1}{\Xp}\right)-
	\left(\Xm-\frac{1}{\Xm}\right)\right]\label{DefE}\,,\\
	Q(X^{\pm})&=&\frac{g}{i}\left[\left(\Xp+\frac{1}{\Xp}\right)-
\left(\Xm+\frac{1}
	{\Xm}\right)\right]\label{DefQ}\,.
\end{eqnarray}
Real values of $E$ and $P$ are obtained by imposing $\Xm=(\Xp)^{*}$.
In the following we will use lower case letters $x^{\pm}$ and
$y^{\pm}$ to denote the spectral parameters in the case of the
elementary magnon $Q=1$.
\paragraph{}
It will be of interest to understand the semiclassical string limit, $g\to
\infty$ of the elementary magnons and their bound states. Importantly,
there are several distinct ways in which the limit can be taken. The
first, which we will call the ``plane-wave'' limit
\footnote{This limit takes its name from its relation to the Penrose
  limit where the dual string
  background becomes a gravitational plane-wave. In the following we
 will see that the terminology is also appropriate for an unrelated 
reason, namely that the magnon is naturally associated with the plane-wave
  solutions of the linearized equation of motion in this limit.}
\cite{BMN} is given by:
\begin{equation}
	g\to\infty\,,~~~p\sim \frac{1}{g}\,,~~~Q~~{\rm Fixed}\,.\label{PPlimit}
\end{equation}
In terms of the spectral parameters $X^{\pm}$, this can be
equivalently imposed by 
\begin{equation}
	\Xp\sim\Xm\approx r+\cO(1/g)\,,~~~~ r \in \bbR \label{PPlimit2}\,,
\end{equation}
so that the dispersion relation for the magnon and its bound states becomes
\begin{equation}
	\Delta-J=\sqrt{Q^{2}+k^{2}}\,,\label{PPDisp}
\end{equation}
where the combination $k=2gp\sim \cO(g^{0})$ is kept fixed in the
limit (\ref{PPlimit2}). In this limit, the magnon goes over to the
elementary excitation of the worldsheet fields of the $AdS_{5}\times
S^{5}$ string. In canonical quantisation these states are the quanta
associated with linearised fluctuations of the worldsheet fields
around a point-like string (the BMN solution \cite{BMN}) which orbits
the equator
of $S^{5}$ at the speed of light. The fluctuations take the form of
plane-waves which solve the linearised equations of motion of the
worldsheet theory and have the form,    
\begin{equation}
	\delta Z(x,t) \sim \exp(i\omega t+ik x)\label{PPWF}\,,
\end{equation}
where $Z$ is a complex worldsheet field.  The frequency
$\omega=\sqrt{1+k^{2}}$ and the wave number $k$ can also be written in
term of the magnon spectral parameters $x^{\pm}\simeq r$ as
\begin{equation}
	\omega(r)=\frac{r^{2}+1}{r^{2}-1}\,,~~~k(r)=\frac{2r}{r^{2}-1}\,.
	\label{erkr}
\end{equation}
States with $Q>1$ correspond to bound states of the elementary
worldsheet excitations in this limit. 
\paragraph{}
A second interesting limit is the so-called
``Giant Magnon'' limit \cite{HM} which  corresponds to
\begin{equation}
	g\to \infty\,,~~~p~~{\rm Fixed}\,;\label{GMlimit1}
\end{equation}
for a BPS state of fixed charge $Q$.  Equivalently, in terms of the
corresponding spectral parameters we have, 
\begin{equation}
	\Xp\sim \frac{1}{\Xm}\approx \exp(ip/2)+\cO(1/g)\,.\label{GMlimit2}
\end{equation}
In this limit the spin-chain magnon 
and its bound states correspond to a classical soliton
configuration on the string worldsheet with dispersion,
\begin{equation}
	\Delta-J\approx
	4g\left|\sin\left(\frac{p}{2}\right)\right|\,\,\,+\,\,\,\
	O(1/g)
	\label{GMDisp}\,.
\end{equation}
The corresponding
string energy $E=\Delta-J$ scales linearly in $g$ as appropriate for a
classical soliton. In the target spacetime, 
the worldsheet soliton is identified with loop of
open string with endpoints on an equator of $S^{5}$. 
It is interesting to note that there is no
$\cO(g^{0})$ term in the expansion of the exact dispersion relation
(\ref{MagDisp}) in this limit.  This indicates that the soliton energy
does not receive a correction at one-loop order in the
semiclassical expansion, which corresponds to an expansion in powers
of $1/g$.
\paragraph{}
Although the magnon looks quite different in the plane-wave and giant
magnon limits described above, it is possible to smoothly interpolate
between the two cases. The elementary quantum of the worldsheet theory
and the classical soliton are representatives of the same excitation
in different regions of momentum space.  This is particularly clear if
one considers the ``near flat-space'' limit  introduced \cite{MalSwan}
where $\xp\sim \xm\sim 1$. 
\paragraph{}
The Giant Magnon limit discussed above is identical for all BPS states
of fixed charge $Q$. As the charge is an adjustable parameter we can
also take a different limit,  
\begin{equation}
	g\to\infty\,,~~~ Q\sim g\,,~~~p~~{\rm Fixed}\,;\label{DGMlimit}
\end{equation}
where the spectral parameters $X^{\pm}$ remain fixed and, as before, 
obey the constraint:
\begin{equation}
	\left(\Xp+\frac{1}{\Xp}\right)-\left(\Xm+\frac{1}{\Xm}\right)=
	i\frac{Q}{g}\sim \cO(g^{0})\label{Qconstraint}\,.
\end{equation}
This limit yields a family of classical soliton
configurations of the worldsheet theory, with energy
\begin{equation}
	\Delta-J=\sqrt{Q^{2}+16g^{2}\sin^{2}\left(\frac{p}{2}\right)}
	\sim\cO(g)\label{DGMDisp}\,,
\end{equation}
where $Q\sim g$ is now regarded as a continuous
parameter\footnote{Like all the soliton solutions considered here the
  solution also depends non-trivially on the momentum $p$ and the
  initial position $x^{(0)}$.}. These
solutions are known as ``Dyonic Giant Magnons'' (DGMs) \cite{CDO1}
(see also \cite{Dressing, AFZ, Minahan1}) and we
will refer to the corresponding limit as the ``DGM limit''.  
The previously discussed Giant Magnon of \cite{HM} is obtained as a 
smooth $Q\to 0$ of
this more general solution. As the DGM dispersion relation coincides
with the exact dispersion relation ({\ref{MagDisp}), the only
correction is the quantisation of the charge $Q$ integer units. 
As in the ordinary
Giant Magnon case, we should therefore expect that the one-loop
correction to the soliton energy vanishes. In the following we will
check this vanishing by a direct calculation in the worldsheet theory.  

\subsection{The Magnon Scattering Matrix}\label{MSmatrix}
\paragraph{}
The exact S-matrix for two elementary magnons in the same $\fsu(2)$
sub-sector takes the form,
\begin{equation}
	\is_{\fsu(2)}(x^{\pm},y^{\pm})=
\is_{\rm BDS}(x^{\pm},y^{\pm})\sigma^{2}(x^{\pm},y^{\pm})\,,~~~~~
	\is_{\rm
		BDS}(x^{\pm},y^{\pm})=
\frac{\xp-\ym}{\xm-\yp}\frac{1-1/\xp\ym}{1-1/\xm\yp}
	\label{fullsu2Smatrix}\,.
\end{equation}
Here the factor $\is_{\rm BDS}(x^{\pm},y^{\pm})$ originates in the conjectured
all-loop Bethe Ansatz of \cite{BDS}, and
$\sigma(x^{\pm},y^{\pm})=\exp(i\theta(x^{\pm},y^{\pm}))$ is known as 
the ``dressing factor''and  $\theta(x^{\pm},y^{\pm})$ 
will be called the ``dressing phase''. An
exact form for the dressing phase was recently conjectured in
\cite{BES}. Following earlier important work \cite{BHL}, the authors
implemented the constraints of crossing symmetry \cite{Janik} and
Kotikov-Lipatov's principle of maximal transcendentality to obtain an explicit
expression for the phase \cite{KotLip, KotLip2} (see also \cite{ES}
for earlier proposal using transcendentality principle). The poles of the
resulting magnon S-matrix correspond precisely with expectations based
on the exact spectrum (\ref{MagDisp}) \cite{DHM}.   
In the strong coupling, expansion, the
conjectured phase of \cite{BES} reproduces the previously obtained
tree-level \cite{AFS} and the one-loop \cite{HL} contributions. 
\paragraph{}
The exact dressing phase $\theta(x^{\pm},y^{\pm})$ is anti-symmetric
under the interchanges of spectral parameters and can be written as, 
\begin{equation}
	\theta(x^{\pm},y^{\pm})=k(\xp,\yp)-k(\xp,\ym)-k(\xm,\yp)+k(\xm,\ym)
	\label{antiSum}\,.
\end{equation}
In the strong-coupling limit, $g\rightarrow\infty$,
$\theta(x^{\pm},y^{\pm})$ and $k(x,y)$ can be expanded as,
\begin{eqnarray}
	\theta(x^{\pm},y^{\pm})&=&g\theta_{0}(x^\pm,y^\pm)+\theta_{1}(x^\pm,y^\pm)+\cO(1/g)
	\label{thetaexpansion}\,,\\
	k(x,y)&=&gk_{0}(x,y)+k_{1}(x,y)+\cO(1/g)\label{Kexpansion}\,.
\end{eqnarray}
respectively.  The explicit form of the tree-level contribution was
first proposed in \cite{AFS} and take the form, 
\begin{equation}
	k_{0}(x,y)=\left[\left(y+\frac{1}{y}\right)-\left(x+\frac{1}{x}\right)
	\right]\log\left(1-\frac{1}{xy}\right)\label{Defk0}\,.
\end{equation}
The one-loop term $k_{1}(x,y)$ was first obtained in \cite{HL} from
considering the quantum fluctuations certain spinning string solution
and can be written as \cite{AF}, 
\begin{eqnarray}
	k_{1}(x,y)&=&\kappa_{1}(x,y)-\kappa_{1}(y,x)\,,\label{Defk1}\\
	\kappa_{1}(x,y)&=&\frac{1}{\pi}\log\left(\frac{y-1}{y+1}\right)
	\log\left(\frac{x-1/y}{x-y}\right)\nn\\
	&+&\frac{1}{\pi}\left[\Li2\left(\frac{\sy-1/\sy}{\sy-\sx}\right)
	-\Li2\left(\frac{\sy+1/\sy}{\sy-\sx}\right)
	+\Li2\left(\frac{\sy-1/\sy}{\sy+\sx}\right)-
	\Li2\left(\frac{\sy+1/\sy}{\sy+\sx}\right)\right]\,.\nn\\
	\label{defchi1}
\end{eqnarray}
\paragraph{}
In the following our main concern will be with the consequences of the
above expressions for the semiclassical scattering of worldsheet
solitons. In particular, in the limit $g\rightarrow\infty$, the
expression (\ref{Defk0}) determines the leading semiclassical
contribution to the S-matrix of two Giant Magnons. This prediction was
checked against a first-principles calculation in \cite{HM}. One of
the main aims of this paper is to extend this check to the next order
in $1/g$. In this regard, it is important to note that the
correspondence between the expansions (\ref{thetaexpansion}) and
(\ref{Kexpansion}) and the semiclassical expansion of the worldsheet
theory is not quite straightforward. The reason is that the magnon
spectral parameters contain hidden dependence on the coupling $g$
because of the constraint, 
\begin{eqnarray}
	\left(x^{+}+\frac{1}{x^{+}}\right) -\left(x^{-}+\frac{1}{x^{-}}\right)
	&= & \frac{i}{g}
\end{eqnarray}
which follows from (\ref{DefQ}) with $Q=1$. This problem is easily
avoided by working in the slightly more general context of the
scattering of two magnon bound states of charges $Q_{1}$ and $Q_{2}$.
As discussed in \cite{CDO2}, the exact bound state S-matrix can be
constructed from the exact magnon S-matrix via the standard fusion
procedure. The result is conveniently expressed in terms of the
bound state  spectral parameters introduced above as, 
\begin{equation}
	\iS_{\fsu(2)}(X^{\pm},Y^{\pm})=\iS_{\rm BDS}(X^{\pm},Y^{\pm})
	\sigma^{2}(X^{\pm},Y^{\pm})\label{su2BSSmatrix}\,.
\end{equation}
Here $\iS_{\rm BDS}(X,Y)$ is the exact expression constructed from
applying the fusion procedure to the BDS part, $\is_{\rm BDS}(x,y)$,
of the magnon S-matrix in (\ref{fullsu2Smatrix}). The explicit
expression, which will not be needed here, can be found in
\cite{CDO2}. Importantly,  the factor $\sigma^{2}(X^\pm,Y^\pm)$ is exactly the
{\em same} dressing factor appearing in the elementary magnon S-matrix
(\ref{su2BSSmatrix}), the only difference being that the bound state
spectral parameters $X^{\pm}, Y^{\pm}$ replace the spectral parameters
$x^{\pm}, y^{\pm}$ of the fundamental magnons. 
\paragraph{}
We can now take the DGM limit (\ref{DGMlimit}) for both magnon
bound states. As the spectral parameters remain fixed in this limit, the
terms in the strong-coupling expansion of the dressing phase
$\sigma(X^{\pm},Y^{\pm})$ correspond directly to terms in the
semiclassical expansion of the worldsheet theory. The resulting
semiclassical S-matrix can be written in the first two orders as,   
\begin{eqnarray}
	\iS_{\fsu(2)}(X^{\pm},Y^{\pm})&\cong& \exp(2i\Theta(X^{\pm},Y^{\pm}))
	\,,\label{approxsu2BSSmatrix}\\
	\Theta(X^{\pm},Y^{\pm})&=&K(\Xp,\Yp)-K(\Xp,\Ym)-K(\Xm,\Yp)+K(\Xm,\Ym)
	\label{DefTheta}\,,\\
	K(X,Y)&=&gK_{0}(X,Y)+K_{1}(X,Y)+\cO(1/g)\label{expK}\,.
\end{eqnarray}
The function $K_{0}(X,Y)$ was calculated in \cite{CDO2} and checked
against a leading order semiclassical calculation of the dyonic giant
magnon scattering matrix. It is given by
\begin{equation}
	K_{0}(X,Y)=\left[\left(X+\frac{1}{X}\right)-
	\left(Y+\frac{1}{Y}\right)\right]\log(X-Y)\label{DefK0}\,.
\end{equation}
Notice that $K_{0}(X,Y)$ is functionally different from $k_{0}(x,y)$
in (\ref{Defk0}). As explained in \cite{CDO2}, this is due to a
non-trivial contribution from the BDS piece $\iS_{\rm BDS}$. At the
next order, we have
\begin{equation}
	K_{1}(X,Y)=k_{1}(X,Y)\,,
\label{Defk1b}
\end{equation}
where the function $k_{1}$ is defined in (\ref{Defk1}). In other words
the one-loop contribution to the bound state S-matrix comes purely from
the dressing phase and is therefore functionally identical to the
Hernandez-Lopez contribution to the magnon dressing phase.  This can
be traced to the fact that the BDS term $\iS_{\rm
BDS}(X^{\pm},Y^{\pm})$ is analytic in $g^{2}$ and therefore only
contributes at even loop order in the worldsheet expansion. 
\paragraph{}
The main conclusion of this Section concerns the predictions for the
one-loop contributions to the dispersion relation and scattering
matrix of Dyonic Giant Magnons. Specifically we have seen that the
known exact dispersion relation requires that the one-loop correction
to the soliton mass vanishes exactly. The one-loop correction to the
S-matrix can be expressed in terms of the Dyonic Giant Magnon spectral
parameters $X^{\pm}$ and $Y^{\pm}$ defined above and is functionally
identical to the Hernandez-Lopez contribution to the magnon dressing
phase. In the rest of the paper we will test these results against
direct semiclassical calculations. 

\section{Determining the Classical Phase Shifts}
\label{PhaseShifts}
\paragraph{}
As explained in the introduction, the main ingredient in the
calculation of one-loop quantum corrections is the classical
scattering data for small fluctuations around the soliton solution.
In particular we need to determine the phase shifts for classical
plane waves scattering off multiple solitons. In this Section we will
address this problem using three different approaches, each of which
will yield part of the information we need.   
\paragraph{}
The starting point is the Metsaev-Tseytlin action for the
Green-Schwarz superstring in $AdS_{5}\times S^{5}$ in conformal gauge
\cite{MetTsey,MetTsey1e2, MetTsey2}. Here the global embedding that 
parametrises the $AdS_5\times S^5$ spacetime can be chosen as 
\begin{eqnarray}
{AdS_5}&:&\quad -|Y_1|^2+|Y_2|^2+|Y_3|^2=-1\,,\nn\\
{S^5}&:&\quad |Z_1|^2+|Z_2|^2+|Z_3|^2=1\,.
\end{eqnarray}  
For our study of worldsheet scattering matrix, $Y_{1}$ and $Z_1$ are
gauge-fixed to form the longitudinal light-cone coordinates, whereas
$\{Y_2, \bar{Y}_2, Y_3, \bar{Y}_3; Z_2, \bar{Z}_2, Z_3, \bar{Z}_3\}$
become eight bosonic transverse excitations and combine to transform
in the (bi-)vector representation under the residual $SO(4)\times
SO(4)$ subgroup. Similarly for the worldsheet fermions, under such
gauge choice, 
the remaining components (after fixing $\kappa$-symmetry) become        
$\{\theta_1,\dots,\theta_4,\eta_1,\dots,\eta_4\}$,
they combine to transform in the bi-spinor representation of 
$SO(4)\times SO(4)$. 
Together, the eight bosonic and eight fermionic fluctuations form the
bi-fundamental representations of 
residual $PSU(2|2)^2$ symmetry group.
We will consider these sixteen fluctuations around a classical soliton
background, and we shall use a uniform notation to denote them:
\[
\mathcal{I}\equiv\{\overbrace{Y_2,\bar Y_2,Y_3,\bar
Y_3}^{\mathcal{I}_{AdS_5}};\overbrace{Z_2,\bar Z_2,Z_3,\bar
Z_3}^{\mathcal{I}_{S^5}};\underbrace{\theta_1,\theta_2,\theta_3,\theta_4;
\eta_1,\eta_2,\eta_3,\eta_4}_{\mathcal{I}_\text{fermions}}\}.
\] 
\paragraph{}
As we review below, the Dyonic Giant Magnon (DGMs) is a soliton solution 
of the worldsheet
theory for which the corresponding string motion occurs on an
$\mathbb{R}\times S^{3}$ submanifold of $AdS_{5}\times S^{5}$. 
We need to consider linearised fluctuations of all of the world 
sheet fields around the classical solution corresponding to one or
more DGMs. The necessary phase shifts are then encoded in the asymptotics 
of the fluctuations in the limits $x\rightarrow \pm \infty$ where $x$ is 
the space-like worldsheet coordinate. 
In the next subsection we will proceed by constructing multi-DGM
solutions and their classical fluctuation spectra explicitly using the
dressing method. In its present form this method is only applicable to
the bosonic worldsheet fields. In subsection 3.2, we employ a
different method based on the finite gap construction of \cite{KMMZ} 
which also yields the phase shifts for the fermionic worldsheet
fields. Finally, in subsection 3.3, we describe a third method using the
proposed all-loop magnon scattering matrix \cite{BeisertSU22} which
provides further non-trivial checks on our results.     

\subsection{Phase Shifts from the Dressing Method}\label{Dress}
\paragraph{}
In this section we present the semiclassical phase shifts calculated
directly from string sigma model using the so-called {\it ``dressing
method''}.  This is a standard technique for constructing
multi-soliton solutions in classical integrable systems which was
applied in the present context by Spradlin and Volovich
\cite{Dressing}.  
\paragraph{}
As discussed in the previous section, the Dyonic Giant Magnon (DGM) is a
soliton on the string worldsheet \cite{CDO1}. It corresponds to a family of
classical solutions labelled by the conserved momentum $p$ and charge
$Q=J_2$, where $J_2$ is one of the three generators $J_1,\, J_2,\,
J_3$ for the global symmetry group $SO(6)$ of the sphere $S^5$. This
data can be equivalently given by two complex spectral parameters $X^{+}$ and
$X^{-}=(X^{+})^{*}$. The solution is also labelled by its initial
position $x^{(0)}$ as well as some extra parameters which determine
its orientation inside $S^{5}$ at time $t=0$. 
As discussed above, the DGM admits a special limit
where $X^{+}\simeq 1/X^{-}$,  the charge vanishes and the solution
reduces to an ordinary Giant Magnon of the type considered by Hofman
and Maldacena \cite{HM}.  It also admits a limit where $X^{+}\simeq
X^{-}\simeq r$ and it collapses to the vacuum. In the target space the
vacuum configuration is just the BMN string solution describing a
pointlike string orbiting an equator of $S^{5}$. Near this degenerate
point the soliton solution reduces to a solution of the linearised
equations of motion corresponding to a plane wave of small amplitude
with wave number and frequency, 
\begin{equation}
\omega(r)=\frac{r^{2}+1}{r^{2}-1}\,,~~~k(r)=\frac{2r}{r^{2}-1}\,.
\label{erkr2} 
\end{equation} 
\paragraph{} 
As we shall review below, the dressing method allows us to construct
exact multi-soliton solutions of the worldsheet theory.  In particular
we can construct a configuration containing $N$ DGMs with individual
spectral parameters $X^{\pm}_{i}$, for $i=1,2,\ldots,N$. We can now
take a limit where, for example for the $n$-th DGM 
$X^{+}_{n}\simeq X^{-}_{n}$ and the
solution collapses to the one describing $N-1$ DGMs. Near this limit the
exact solution must go over to a solution of the equations linearised
around the $N-1$ soliton solution. As first noted by Dashen,
Hasslacher and Neveu \cite{Dashen}, this construction provides an
elegant way of extracting the exact spectrum of small fluctuations
and, in particular, the corresponding phase shifts. We will now apply
this methodology to the bosonic sector of the worldsheet $\sigma$
model. Some of the the calculation details are relegated to 
Appendix \ref{dressingup}.  
\paragraph{} 
The Dyonic Giant Magnon corresponds to
string motion in an $\mathbb{R}\times S^{3}$ subspace of the the full
$AdS_{5}\times S^{5}$ spacetime. It is easy to check that fluctuations in the
$AdS_{5}$ directions couple trivially to this 
background and thus have vanishing
phase shifts. Thus we will focus on the $S^{5}$ sector of the
worldsheet theory.  Following \cite{Dressing} we work in static gauge 
and the worldsheet theory
in this sector essentially reduces to a bosonic
sigma-model on a flat two-dimensional worldsheet with the coset
${\rm{SU}}(4)/{\rm{Sp}}(2)\approx S^5$ as target space. The equations
of motion of this $\sigma$-model must be also supplemented by
the Virasoro constraints. 
\paragraph{}
The coset construction exploits the existence of a
$\mathbb{Z}_2$-automorphism $\Omega\in\mathrm{Aut}(\SU(4))$, where
\begin{equation}
	\Omega(\g):=J^{-1}\g^T J,\quad\text{where}\quad \g\in {\rm SU}(4)\quad
	\text{and}\quad
	J=\begin{pmatrix} 0 & 0 & 1 & 0 \\ 0 & 0 & 0 & 1 \\
	 -1 & 0 & 0 & 0 \\ 0 & -1 & 0 & 0\end{pmatrix}\,.\label{DefJ}
\end{equation}
It has the property $\Omega^2=1$. The condition
\begin{equation}\label{cosetcond}
	\Omega(\mathcal{P})=\mathcal{P}\quad\text{for}
\quad\mathcal{P}\subset \SU(4)
\end{equation}
will give us $\mathcal{P}\approx \SU(4)/\Sp(2)\approx S^5$. This allows for
a decomposition of $\SU(4)$ into
\[\SU(4)\approx \Sp(2)\otimes \SU(4)/\Sp(2)\,.\]
A convenient parametrisation of the coset is given by, 
\[\g=\begin{pmatrix} Z_1 & Z_2 & 0 & Z_3 \\ - \bar Z_2 & \bar Z_1 & -\bar
Z_3 & 0 \\ 0 & \bar Z_3 & Z_1 & -\bar Z_2 \\ -\bar Z_3 & 0 & Z_2 & \bar
Z_1 \end{pmatrix}\,,\]\label{Defbfg}
where the components $Z_1, Z_2$ and $Z_3$ satisfy
$\sum_{i=1}^{3}|Z_{i}|^{2}=1$. 
By defining the flat current
$j=-\g^{-1}d\g \in \fsu(4)$, we can make the following decomposition,
\[ j=H+P,\quad H\in \fsp(2),\quad P\in \fsu(4)/\fsp(2).\]
\paragraph{}
The equations of motion for the sigma model can then be written
succinctly as
\[d\star P=\star P\wedge H+H\wedge\star P\,,\]
where $\star$ denotes the Hodge-dual with respect to the worldsheet
metric.  These can be equivalently be expressed as the zero curvature
condition of the following flat Lax connection
\[\hat{j}( X)=H+\frac{1+ X^2}{1- X^2}P+\frac{2 X}{1- X^2} \star P,\]
with $X\in\mathbb{C}$ being a spectral parameter, and notice that
$\hat{j}(X=0)=j$. By picking the coordinates $z_{\pm}=\frac12 (x\pm t)$ 
as coordinates in the worldsheet, we find this connection has the form 
\[\hat{j}( X)=H+\frac{\partial_- \g \,\g^{-1}}{1- X}+\frac{\partial_+ \g
\,\g^{-1}}{1+ X}.  \]
The flatness condition for $\hat{j}$ is equivalent to the consistency
conditions for the auxiliary linear problem,
\[\Bigl[\partial_--\frac{\partial_- \g \,\g^{-1}}{1- X}\Bigr]\Psi( X)=0\,,\]
\[\Bigl[\partial_+-\frac{\partial_+ \g \,\g^{-1}}{1+ X}\Bigr]\Psi( X)=0\,.\]
Clearly $\Psi( X=0)=\g$ will be a solution to these equations. Only
those solutions that further obey \eqref{cosetcond} and the Virasoro
constrains will be solutions of the string equations of motion. 
\paragraph{}
The trivial vacuum solution of the equations of motion corresponding to
the BMN point like string solution is given by,  
\[\Psi_0( X)=\mathrm{diag}(e^{iZ( X)},e^{-iZ( X)},e^{iZ( X)},e^{-iZ( X)}),\]
where
\[Z( X)=\frac{z_{-}}{ X-1}+\frac{z_{+}}{ X+1}\,,\]
This solution has vanishing energy
$\Delta-J_1=0$. 
\paragraph{}
The dressing method proceeds by applying a spectral-parameter
dependent gauge transformation to both the connection $\hat j(X)$ and
the auxiliary wave function $\Psi(X)$. It is a solution generating
technique that can be used to map trivial solutions of the equations
of motion into new non-trivial solutions.  Here we review the
construction given in \cite{Dressing, Dressing1}, more details can be found in
\cite{Dressing, Dressing1, Babelon}.  Explicitly, a new solution can be
determined from the vacuum solution by acting on it with a gauge
transformation $\chi_1(X)$,
\[\begin{cases}
\Psi_1( X)=\chi_1(X)\Psi_0( X)\,,\\
\hat j_1(X)=\chi_1(X)\,\hat
j_0(X)\,\chi_1^{-1}(X)+d\chi_1(X)\,\chi_1^{-1}(X)\,,
\end{cases} \]
where $\hat j_0(X)\equiv \hat j(X)|_{\g=\g_0}$ and
\[\chi_1(X)=1+\frac{ X_1- \bar{X}_1}{ X- X_1}\mathcal{P}_1[w_1]+
\frac{1/\bar{X}_{1}-1/ X_1}{ X-1/\bar{X}_{1}}\mathcal{Q}_1[w_1]\,.\]
The projection operators $\mathcal{P}_1,\mathcal{Q}_1$ are determined
from $\Psi_0$ itself by requiring that the dressing transformation
does not change the analytic structure of the Lax connection $\hat
j(X)$ and that $\Psi_1(X)$ obeys \eqref{cosetcond}. 
Here $w_1$ is a four-component vector specifying the orientation of
the solution in the target space. In particular, by taking
$\g_1=\Psi_1(0)$ and making the identifications $X_1=r_1
e^{ip_1/2}\equiv X_1^+$ and $\bar X_1=r_1 e^{-ip_1/2}\equiv X_1^-$,
and selecting the polarisation vector $w_1\equiv
w_\parallel=(1,i,0,0)^t$, we recover the familiar DGM solution of
\cite{CDO1},
\[\g_1=\g_0-\frac{\Xp_1-\Xm_1}{\Xp_1} \Bigl(\mathcal{P}_1[w_1]+
\mathcal{Q}_1[w_1]\Bigr)\g_0\,.\]
This solution has the following conserved quantities:
\begin{align}
	\Delta-J_1&=2g\frac{1+r_1^2}{r_1}\Big|\sin\Bigl(\frac{p_1}2\Bigr)
	\Big|\,, \\
	J_2&=2g\frac{1-r_1^2}{r_1}\Big|\cos\Bigl(\frac{p_1}2\Bigr)
	\Big|\,, \\
	J_3&=0\,.
\end{align}
\paragraph{}
The orientation vector $w_1=w_\parallel$ determines which $SU(2)\simeq
S^{3}$ subspace of the target $S^{5}$ in which the DGM is embedded.
Picking an orthogonal orientation vector $w_1\equiv
w_\perp=(i,0,0,1)^t$ simply has the effect of interchanging the Cartan
charges $J_2$ and $J_3$ and selecting a different $SU(2)$ subspace for
the embedding.   
\paragraph{}
As mentioned above, in the limit
\[p_1\rightarrow 0,\quad r_1\, \text{fixed}\,\Leftrightarrow\,
X_1^+\sim X_1^-\]
the DGM solution goes over to the vacuum, $\g_1(x,t)\rightarrow
\g_0(x,t)$. Expanding $\g_1$ in $\eta\equiv \Xp_1-\Xm_1$, we find
that, at linear order in $\eta$, the resulting solution describes a
plane wave propagating in the background described by $\g_0$. The
dressing method allows us to determine easily an explicit expression
for the perturbed solution by evaluating,
\[\g_1=\g_0+\delta \g_0,\]
with
\[\delta \g_0=-2i\sin\Bigl(\frac{p_1}2\Bigr)\Bigl(\mathcal{P}_1[w_1]+
\mathcal{Q}_1[w_1]\Bigr)\Big|_{\eta=0}\g_0,\]
being the plane-wave solution. For the orientation $w_1\equiv
w_\parallel$ we find,
\begin{eqnarray}
	\delta \g_0&\equiv&\begin{pmatrix} \delta Z_1 & \delta Z_2 & 0 & 
	\delta Z_3 \\
	- \delta \bar Z_2 & \delta \bar Z_1 & - \delta \bar
	Z_3 & 0 \\ 0 & \delta \bar Z_3 & \delta Z_1 & - \delta \bar Z_2 \\
	-\delta \bar Z_3 & 0 & \delta Z_2 & \delta \bar
	Z_1 \end{pmatrix}\\
	&=&-2i\sin\Bigl(\frac{p_1}2\Bigr)\cdot\frac12\begin{pmatrix}
	1 & ie^{iv_1} & 0 & 0 \\
	-ie^{iv_1} & 1 & 0 & 0\\
	0 & 0 & 1 & ie^{iv_1'}\\
	0 & 0 & -ie^{iv_1'} & 1
	\end{pmatrix}
	\begin{pmatrix}
	e^{it} & 0 & 0 & 0 \\
	0 & e^{-it} & 0 & 0 \\
	0 & 0 & e^{it} & 0 \\
	0 & 0 & 0 & e^{-it}
	\end{pmatrix},
\end{eqnarray}
where $v_1\equiv Z(r_1)+\bar Z(r_1)=2Z(r_1)$ and $v'_1=v_1(1/r_1)$.
We then obtain,
\begin{align}
	\delta Z_1&=-i\sin\Bigl(\frac{p_1}{2}\Bigr)\, e^{+it}\,.\\
	\delta Z_2&=\sin\Bigl(\frac{p_1}{2}\Bigr)\, e^{i\omega_1
	t-ik_1x}\,,\\
	\delta Z_3&=0\,.
\end{align}
Thus the perturbation has the form of a plane wave with wave number
given by $k_1=2r_1/1-r_1^2$ and frequency
$\omega_1=1+r_1^2/1-r_1^2=\sqrt{k^2_1+1}$. As the background is the
trivial vacuum there is no phase shift.  We can also take an
orthogonal orientation vector $w_{1}=w_\perp$ to obtain identical
results but with $\delta Z_2$ and $\delta Z_3$ interchanged. 
\paragraph{}
We can now apply the same technique to determine the solution
describing the propagation of a plane-wave in the $n$-soliton
background.  Since we merely need to determine the phase shifts
$\delta_{Z_k}(r;\{X_j^\pm\})$ and $\delta_{\bar Z_k}(r;\{X_j^\pm\})$
corresponding to the fields $\delta Z_k$ and $\delta \bar Z_k$, we
will be only interested in the asymptotic limits of this perturbation
solution rather than the full solution. The phase shifts in general
can then be calculated from:
\begin{equation}
	\delta_{Z_k}(r;\{X_j^\pm\})=-i\log\left(\delta Z_k\right)\Big|_{+\infty}
	-i\log\left(\delta Z_k\right)\Big|_{-\infty}\,.
\end{equation}
Here we only list the results calculated from this approach, and we
present the relevant calculation details in the Appendix
\ref{dressingup}.  The polarisations within this sector will be
labelled by the coordinates that suffer a non-trivial phase shift,
$I\in \mathcal{I}_{S^5}\equiv\{Z_2,\bar Z_2, Z_3,\bar Z_3\}$, i.e., a plane-wave aligned
with the background soliton will have a non-trivial phase-shift in the
directions $Z_2,\bar Z_2$, whether a plane-wave with a perpendicular
polarisation will have a phase shift for $Z_3,\bar Z_3$.
\begin{align}
	\delta_{Z_2}\left(r;\{ X_j^\pm\}\right)&=-\delta_{\bar
	Z_2}\left(1/r;\{X_j^{\pm}\}\right)=-2i\sum_{j=1}^N\log\left(\frac{r-
	X_j^+}{r- X_j^-}\right)-P,\label{dZ2}\\
	\delta_{Z_3}\left(r;\{X_j^{\pm}\}\right)&= \delta_{\bar
	Z_3}\left(r;\{X_j^{\pm}\}\right) =-i\sum_{j=1}^N\log\left(\frac{r-
	X_j^+}{r-X_j^-}\right) -i\sum_{j=1}^{N}\log\left(\frac{1/r-
	X_j^-}{1/r- X_j^+}\right),\label{dmZ3}
\end{align}
where $P\equiv\sum_{j=1}^N p_j$ is the total dyonic giant magnon
momentum and $r=r(k)$ is related to the plane-wave momentum
$k$ by,
\begin{equation}\label{kfuncr}
	k=\frac{2r}{r^2-1}.
\end{equation}
In the GM limit $X_j^{\pm}\rightarrow x^\pm_j\equiv \exp(\pm i p_j/2)$
the phase shifts  take the form,
\begin{equation}
	\delta_{I}(r;\{x^\pm_j\})=
	-2i\sum_{j=1}^N\log\left(\frac{
	r-x^+_j}{r-x^-_j}\right)-P,\quad I\in\mathcal{I}_{S^5}.
\end{equation}
\paragraph{}
Although the dressing method can not be directly applied to the
fermionic case, a fermionic solution for a single Giant Magnon background
was presented in \cite{Spradlin} (See also earlier results in
\cite{Minahan0}).  From there one easily determines the phase shift
for the fermionic perturbations around  an one-giant magnon soliton
background with momentum $p=-i\log(x^+/x^-)$ as
\begin{equation}\label{dmFerm}
	\delta_{I}(r;x^\pm)=-i\log\Bigl(\frac{r-x^+}{r-x^-}\Bigr)-\frac{p}{2},\quad
	I\in \mathcal{I}_{\text{fermions}}\equiv\{\theta_1,\dots,\theta_4;\eta_1,\dots,\eta_4\}.
\end{equation}
As dictated by supersymmetry, the dispersion relation for a 
fermionic perturbation 
is identical to that of the bosons; 
$\omega=\sqrt{k^2+1}$ \cite{Dressing}, with $k$ the plane-wave
momentum, related to $r$ by
\eqref{kfuncr}.

\subsection{Phase Shifts from Finite-Gap Solutions}\label{FGsolution}
\paragraph{}
In the previous section, by applying the dressing method to the
$S^5$ sector, we were able to determine the phase shift caused by
the scattering between a plane-wave bosonic fluctuation and a
$N$-dyonic giant magnon soliton within certain $S^3\subset S^5$.
Extending the dressing method to the full theory including fermionic
fluctuation remains an unsolved problem. In this subsection we will
sidestep this difficulty by using another formalism \cite{KMMZ} which
allows us to construct 
the spectral data for solutions of the worldsheet $\sigma$-model with 
closed-string boundary conditions. In particular, the
worldsheet fields are now taken to be periodic in the 
spatial coordinate $x$ with period $\ell$. In static gauge, where the
energy density is constant along the string, the period is related to
the string energy as $\ell=\Delta/2g$.    
We will
consider string solutions with large but finite energy. 
Thus, for the moment, we
are moving away from the strict Hofman-Maldacena limit described above
where the string becomes infinitely long. 
For periodic boundary
conditions the spectrum of fluctuations around a given classical
background now becomes discrete. As we review below, 
the classical phase shift naturally appears in the corresponding 
quantisation condition for the wave number of the small fluctuations. 
If we pick a classical background which goes over to the DGM solution 
in the limit $\ell\rightarrow \infty$, 
we can then extract the required phase shifts for each worldsheet
field. 

\subsubsection{Dyonic Giant Magnons as Finite-Gap Solutions}\label{DGMFG}
\paragraph{}
We will begin this subsection by 
reviewing the elegant description of classical solutions with periodic
boundary conditions
obtained in \cite{KMMZ} by Kazakov,
Marshakov, Minahan and Zarembo (KMMZ). To start with we will restrict
our attention to states in a particular $SU(2)$ sector where 
the dual string motion is confined to an $\mathbb{R}\times
S^{3}$ submanifold of the spacetime. As mentioned in the previous
section equations of motion for
the bosonic string admit a Lax formulation, with flat connection $j$,  
which immediately implies the existence of an infinite number of
conserved charges at the classical level. The relevant classical solutions
are naturally classified by the analytic behaviour of the 
corresponding monodromy matrix, $\Omega(X)=P\exp(\oint j)$, 
and its eigenvalues as functions of the complex
spectral parameter $X\in \mathbb{C}$ introduced above. 
For classical strings on $\mathbb{R}\times S^{3}$, 
the monodromy matrix is a unimodular $2\times 2$ matrix with
eigenvalues $\exp(\pm ip(X))$. Here, the quasi-momentum $p(X)$ is a
complex function of the spectral parameter with prescribed 
singularities and asymptotics. In particular, $p(X)$ has 
poles with equal residue $-\Delta/4g$ at the points $X=\pm 1$
and can also have branch-cuts denoted $\mathcal{C}_{k}$ for
$k=1,\ldots, K$. Its 
discontinuity across each cut is fixed by the
equation, 
\begin{equation}
p(X+i\epsilon)+p(X-i\epsilon)=2\pi n_{k} 
\label{disc}
\end{equation}
for all $X\in \mathcal{C}_{k}$. The integer $n_{k}$ associated with
each cut is directly related to the mode number of a corresponding
string oscillator. 
The quasi-momentum is properly defined as an
abelian integral of a meromorphic differential on an appropriate
branched covering of the complex $X$-plane. The behaviour of the 
quasi-momentum at these branch cuts can be encoded by expressing it in
terms of a resolvent $G(X)$ as, 
\begin{equation}  
p(X)=G(X)-\frac{\Delta}{4g}\left[\frac{1}{X-1}+\frac{1}{X+1}\right]
\end{equation}
where the resolvent is defined in terms of a positive density
$i\rho(X)$ which is non-zero along a contour $\mathcal{C}=\mathcal{C}_{1}\cup
\mathcal{C}_{2}\ldots\cup \mathcal{C}_{K}$ 
whose connected component are the branch cuts,  
\begin{equation}
G(X)=\,\int_{C}\, dY \, \frac{\rho(Y)}{X-Y}.
\end{equation} 
From (\ref{disc}), we find that the 
resolvent satisfies the fundamental equation, 
\begin{equation}\label{KMMZ}
	G(X+i\epsilon)+G(X-i\epsilon)\equiv
	2\dashint_{\mathcal{C}}\frac{\rho(Y)}
{X-Y}\,dy=2\pi
	n_k+\frac{\Delta}{2g}\Bigl[\frac{1}{X-1}+\frac1{X+1}\Bigr].
\end{equation}
\paragraph{}
The conserved charges $E=\Delta-J$, $Q$ and worldsheet
momentum $p$ of the classical string solution 
are each determined in terms of the density $\rho(X)$ as, 
\begin{eqnarray}
\int_{\mathcal{C}} \,dX \, \rho(X) & = & \frac{1}{2g}\left(E+Q\right)\,,
\label{E+Q}\\ 
 \int_{\mathcal{C}} \,dX \, \frac{\rho(X)}{X} & = & p\,,
\label{momentum}\\ 
 \int_{\mathcal{C}} \,dX \, \frac{\rho(X)}{X^{2}} & = & \frac{1}{2g}
\left(E-Q\right)\,.
\label{dens}
\end{eqnarray}
In general the allowed configurations of the density $\rho(X)$ are
determined by solving the integral equation (\ref{KMMZ}). This leads to 
families of solutions where $\rho$ varies non-trivially along the square
root branch cuts of $p(X)$. The system also admits another type of
configuration where $\rho(X)$ remains constant along certain contours in
the $x$-plane. This leads
instead to logarithmic branch points of the quasi-momentum. The
corresponding branch cuts are referred to as ``condensate cuts''.   
\paragraph{}
In the present case we are interested in the case of large energy 
$\Delta>>1$. In this case, the square root
branch cuts shrink to zero size and non-trivial configurations are
described by condensate cuts alone. The simplest such configuration
is a single condensate cut with constant density $i\rho(x)=1$ and
endpoints at $X=X^{+}$ and $X=X^{-}$. The corresponding resolvent is, 
\begin{equation}
	G(X;X^\pm)=-i\int^{\Xp}_{\Xm}\frac{dY}{X-Y}=\frac{1}{i}
	\log\left(\frac{X-\Xp}{X-\Xm}\right)\label{DGMresolvent}\,.
\end{equation}
As we explain below, this is the fundamental quantity we need for obtaining the
scattering phase for fluctuations around the dyonic giant magnon.
Applying the relations
(\ref{E+Q}), (\ref{dens}) and (\ref{momentum}), we immediately obtain
respectively the formulae for the conserved charges (\ref{DefE}),
(\ref{DefQ}) and 
(\ref{Defp}). 
We can also eliminate the dependence on the endpoints $X^{\pm}$ in these expressions to obtain the dispersion relation, 
\begin{equation}
E=\sqrt{Q^{2}+16g^{2}\sin^{2}\left(\frac{p}{2}\right)}
\label{disp2}
\end{equation}
This precisely matches the dispersion relation for the Dyonic
Giant Magnon (DGM) solution of classical string theory on 
$\mathbb{R}\times S^{3}$ \cite{CDO2} and 
it is natural to identify the condensate cut configuration described
above as the KMMZ spectral data corresponding to this classical
solution \cite{Minahan1, KORS, benoit}. 
In this classical context, the conserved charge $Q$ is a
continuous parameter. 
The original Giant Magnon solution of Hofman and Maldacena \cite{HM} 
is obtained by taking the limit $Q\rightarrow 0$ of this more general
configuration. 
\paragraph{} 
Now let us consider a perturbation around the dyonic giant magnon
solution with resolvent (\ref{DGMresolvent}) described above.  
In our discussion of the dressing method in the previous subsection,
the fluctuation corresponded to the introduction of an additional
``small'' soliton. The corresponding perturbation of the finite gap 
data is to introduce a single additional pole in the quasi-momentum
$p(X)$ \cite{GV1,GV2}. Roughly speaking this can also be thought of as 
the limiting configuration obtained by shrinking an additional
condensate cut corresponding to an additional DGM. 
To ensure that the new configuration with
the additional simple pole remains a solution to the equations of
motion, the position $X=r\in \mathbb{R}$ of the pole is not arbitrary, but is
determined by the fundamental equation \eqref{KMMZ} which now reads,    
\begin{equation}\label{KMMZ1}
	2G(r;X^\pm)=2\pi \tilde
	n+\frac{\Delta}{2g}\Bigl[\frac{1}{r-1}+\frac1{r+1}\Bigr], \quad
	\tilde n \in\mathbb{Z}.
\end{equation}
The worldsheet momentum associated with the additional pole at $x=r$ is 
simply that of a corresponding plane wave excitation  (\ref{erkr}) 
of wavenumber $k(r)=2r/(r^2-1)=1/(r-1)+1/(r+1)$. 
As mentioned above the length $\ell$, of the corresponding closed 
string (measured in the worldsheet coordinate $x$ which is normalised
to be conjugate to the wavenumber $k$) 
is related to the string energy as $\ell=\Delta/2g$. 
We then obtain the following equation from (\ref{KMMZ1}),
\begin{equation}\label{KMMZ1b}
	2G(r;X^\pm)+k(r)\ell=2\pi \tilde n,\quad\tilde
	n \in\mathbb{Z}.
\end{equation}
This equation is responsible for quantising the allowed
values of the wave-number $k(r)$. One then immediately
recognises the first term on the LHS of the above equation as 
the additional phase-shift acquired by the plane-wave fluctuation 
as it travels a full
period $\ell$ of the string, 
\begin{equation}
	\delta_{Z_2}(r;X^\pm)=2G(r;X^\pm)=-2i\log\Bigl(\frac{r-X^+}{r-X^-}\Bigr).
\label{delpara}
\end{equation}
This precisely matches the result given in the previous subsection 
for the phase shift for excitations inside the $SU(2)$ sector
(see Eqn (\ref{dZ2})) up to an additive constant linearly proportional to the
DGM momentum $p$\footnote{Such additive constants can be attributed to
  the different basis choices between string and gauge theories
  c.f.\cite{AFZ1}, and most importantly such ambiguities do not
  contribute to the calculations of the energy shift and the one-loop
  correction to the scattering phase.}.

\subsubsection{Embedding in full $AdS_5\times S^5$}
\paragraph{}
We will now apply the method described in the previous subsection to
the full $AdS_{5}\times S^{5}$ background to recover 
the phase shifts for the fluctuations of
each worldsheet field in the dyonic giant magnon background (See \cite{GV1,GV2,BeiSak} for earlier work).
The full superstring theory is described by a sigma model that has
coset target space
\[\frac{PSU(2,2|4)}{Sp(2,2)\times Sp(4)},\]
and the Virasoro constraint imposed. An element
$\g\in SU(2,2|4)$ has the following form\footnote{$PSU(2,2|4)$ does
not allow a matrix representation.} 
\[\g=\begin{pmatrix} A & B \\ C & D\end{pmatrix}\,,\]
and the coset can be constructed from the existence of an
$\mathbb{Z}_4$-automorphism
$\Omega\in\mathrm{Aut}(PSU(2,2|4))$ with
\[\Omega(\g):=\begin{pmatrix} EA^TE & -ECE \\ EB^TE &
ED^TE\end{pmatrix}\quad\text{and}\quad E=\begin{pmatrix}
0 & -1 & 0 & 0 \\ 
1 & 0 & 0 & 0 \\
0 & 0 & 0 & -1\\
0 & 0 & 1 & 0
\end{pmatrix}.\]
We can then identify $\mathcal{H}=\Omega(\mathcal{H})$, from which one
gets $\mathcal{H}\approx Sp(2,2)\times Sp(4)$.
This model is classically integrable, and its Lax connection is
\begin{equation}
    \hat J(X)=H+\frac{X^2+1}{X^2-1}P-\frac{2X}{X^2-1}\Bigl(\star
	P-\Lambda\Bigr)+\sqrt{\frac{X+1}{X-1}}Q^1+\sqrt{\frac{X-1}{X+1}} Q^2.
\end{equation}
Its flatness condition reproduces the worldsheet equations of motion
for the IIB superstring on $AdS_{5}\times S^{5}$.
\paragraph{}
A convenient parametrisation for the eigenvalues of the monodromy
matrix is given as follows,
\[\Bigl\{e^{i\hat p_1},e^{i\hat p_2},e^{i\hat p_3},e^{i\hat p_4}|
e^{i\tilde p_1},e^{i\tilde p_2},e^{i\tilde p_3},e^{i\tilde
p_4}\Bigr\}.\] 
The quasi-momenta $\hat p_{1,\dots,4}$ and $\tilde p_{1,\dots,4}$ will
then be meromorphic functions over the spectral curve $\Gamma$. They
will define the 8-sheets of the Riemann surface that will characterise
the solution. These sheets will be connected by a set of cuts
$\mathcal{C}_1,\dots,\mathcal{C}_n$ that define the curve. At these
cuts the quasi-momenta can jump by a multiple of $2\pi$,
\begin{equation}
    p_{i}(X+i\epsilon)-p_{j}(X-i\epsilon)=2\pi n_{ij},\quad X\in\mathcal{C}_k^{ij}\,.
\end{equation}
This equation is the generalisation of Eqn (\ref{disc}) appearing in
the analysis of the previous section. 
\paragraph{}
The monodromy matrix obeys \cite{GV1,GV2} the equation
\[ C^{-1}\Omega(X) C=\Omega^{-ST}(1/X), \quad\text{with}\quad 
C=\begin{pmatrix} E & 0 \\ 0 & -iE\end{pmatrix}\,.\]
This symmetry of the monodromy matrix translates into the following
equations for the quasi-momenta,
\begin{align}
    \tilde p_{1,2}(X)&=-\tilde p_{2,1}(1/X),\label{p12}\\
    \tilde p_{3,4}(X)&=-\tilde p_{4,3}(1/X),\label{p34}\\
    \tilde p_{1,2,3,4}(X)&=-\tilde p_{2,1,4,3}(1/X).\label{p1234}
\end{align}
These will be of ultimate importance in fixing the quasi-momenta
on all the sheets.
\paragraph{}
To determine the spectral curve corresponding to finite gap solution
that giving rise to the dyonic giant magnon, we make use of this
symmetry to embed the $SU(2)$ sector solution in the full theory,
\[\tilde p_2(X)=-\tilde
p_3(X)=p_{SU(2)}(X)=G(X;X^\pm)-\frac{\Delta}{2g}\frac{X}{X^2-1}.\]
From this and the (\ref{p12}) above we obtain,
\[\tilde p_1(X)=-\tilde
p_2\Bigl(1/X\Bigr)=-G\Bigl(1/X;X^\pm\Bigr)+\frac{\Delta}{2g}\frac{1/X}{1/X^2-1}=
-G\Bigl(1/X;X^\pm\Bigr)-\frac{\Delta}{2g}\frac{X}{X^2-1}.\]
Likewise from (\ref{p34}) we obtain
\[\tilde p_4(X)=-\tilde p_3\Bigl(1/X\Bigr)=p_{SU(2)}\Bigl(1/X\Bigr)=
G\Bigl(1/X;X^\pm\Bigr)+\frac{\Delta}{2g}\frac{X}{X^2-1}.\]
Repeating the same procedure we determine the relations
between all quasi-momenta and the $\fsu(2)$ sub-sector resolvent $G(X)$,
\begin{align}
	\tilde p_1(X)&=-\tilde p_4(X)=-G\Bigl(1/X;X^\pm\Bigr)-\frac{\Delta}{2g}\frac{X}{X^2-1},\\
	\tilde p_2(X)&=-\tilde p_3(X)=G(X;X^\pm)-\frac{\Delta}{2g}\frac{X}{X^2-1},\\
%\tilde p_3(X)&=-G(X)+\frac{\Delta}{2g}\frac{X}{X^2-1},\\
%\tilde
%p_4(X)&=G\Bigl(\frac1{X}\Bigr)+\frac{\Delta}{2g}\frac{X}{X^2-1},\\
	\hat p_{1,2}(X)&=-\hat p_{3,4}(X)=-\frac{\Delta}{2g}\frac{X}{X^2-1}.
%\hat p_{3,4}(X)&=\frac{\Delta}{2g}\frac{X}{X^2-1}.
\end{align}
We now apply the same method as before: we introduce a microscopic
probe cut (or, more simply, a pole), corresponding to a small
fluctuation, 
which can connect any of
the eight-sheets. The connection between the excitations of specific
worldsheet fields and cuts connecting particular pairs of sheets of
the spectral curve was given in \cite{GV1}: 
\begin{align}
	 S^5&:\quad (i,j)=\overbrace{(\tilde 1,\tilde
3)}^{Z_3},\,\overbrace{(\tilde 1,\tilde
	4)}^{\bar Z_2},\,\overbrace{(\tilde 2,\tilde
3)}^{Z_2},\,\overbrace{(\tilde 2,\tilde 4)}^{\bar Z_3},\\ AdS_5&: \quad
	(i,j)=\overbrace{(\hat 1,\hat 3),\,(\hat 1,\hat
4),\,(\hat 2,\hat 3),\,(\hat
	2,\hat 4)}^{Y_2,\,\bar Y_2,\,Y_3,\,\bar Y_3},\\ \text{fermionic}&: \quad (i,j)=\overbrace{(\tilde 1,\hat
	3),\,(\hat 1,\tilde 4),\,(\tilde 2,\hat 3),\,(\hat 2,\tilde 4)}^{\eta_1,\, \eta_2,\,\eta_3,\,\eta_4}, \overbrace{(\hat
	1,\tilde 3),\,(\tilde 1,\hat 4),\,(\hat 2,\tilde 3),\,(\tilde
2,\hat 4)}^{\theta_1,\,\theta_2,\,\theta_3,\,\theta_4}.  
\end{align}
So for instance a cut connecting the sheets $\tilde 2$ and $\tilde 3$
will be a perturbation inside $S^{3}\subset S^5$ associated with the
$\fsu(2)$ sub-sector, i.e., it will be a fluctuation with a polarisation along
$Z_2$.  Applying the KMMZ equation to the probe cut we
will then have \[\tilde p_2(r)-\tilde p_3(r)=2\pi n_{23},\quad
n_{23}\in\mathbb{Z},\] that translates, in the language of the
$\fsu(2)$ sector as \[2G(r;X^\pm)-k(r)\ell=2\pi \tilde n,\quad \tilde
n\in\mathbb{Z},\] which coincides with Eqn (\ref{KMMZ1b}). 
\paragraph{}
Repeating this to all other polarisations, we get
for the full $S^5$ sector
\begin{align}
    \tilde p_1(r)-\tilde p_3(r)=2\pi n_{13} &\Rightarrow
G(r;X^\pm)-G(1/r;X^\pm)-k(r)\ell
	=2\pi n_{13},\label{p13}\\
    \tilde p_1(r)-\tilde p_4(r)=2\pi n_{14} &\Rightarrow
	-2G(1/r;X^\pm)-k(r)\ell=2\pi n_{14},\label{p14}\\
    \tilde p_2(r)-\tilde p_3(r)=2\pi n_{23} &\Rightarrow 2G(r;X^\pm)-k(r)\ell=2\pi
	n_{23}\label{p23},\\
    \tilde p_2(r)-\tilde p_4(r)=2\pi n_{24} &\Rightarrow
	G(r;X^\pm)-G(1/r;X^\pm)-k(r)
\ell=2\pi n_{24}.\label{p24}
\end{align}
For the $AdS_5$ sector, these are trivial as expected:
\[k(r)\ell=2\pi n_{13}=2\pi n_{14}=2\pi n_{23}=2\pi n_{24}\,.\]\label{pAdS}
Lastly for the fermions we have,
\begin{eqnarray}
   -G(1/r;X^\pm)-k(r)\ell&=&2\pi n_{\tilde 1\hat 3}=2\pi n_{\tilde 1\hat 4}=
	2\pi n_{\hat 1\tilde 4}=2\pi n_{\hat 2\tilde 4}\,,\label{fermi1}\\
	G(r;X^\pm)-k(r)\ell&=&2\pi n_{\tilde 2\hat 3}=2\pi n_{\tilde 2\hat 4}=2\pi 
	n_{\hat 1\tilde 3}=2\pi n_{\hat 2\tilde 3}\label{fermi2}\,.
\end{eqnarray}
where in all of these equations $G(r;X^\pm)$ is the $SU(2)$ resolvent
for a dyonic giant magnon solution,
\begin{equation}\label{resolventDGM}
	G(r;X^\pm)=\frac1{i}\log\Bigl(\frac{r-\Xp}{r-\Xm}\Bigr),
\end{equation}
and
\[k(r)=\frac{2r}{r^{2}-1}.\]
Comparing these equations (\ref{p13})-(\ref{fermi2}) with the
periodicity equation (\ref{KMMZ1b}) one can immediately read off the
various phase shifts:
\begin{itemize}
\item For the $SU(2)$ or $S^{3}$ sub-sector:
%$\mathcal{I}_{S^3}\equiv\{Z_2,\bar Z_2\}\subset\mathcal{I}$:
    \begin{align}\label{fgZ2}
        \delta_{Z_2}(r;X^\pm)&\equiv\delta_{\tilde 2\tilde
3}(r;X^\pm)=2G(r;X^\pm),\\
        \delta_{\bar Z_2}(r;X^\pm)&\equiv\delta_{\tilde 1\tilde
4}(r;X^\pm)=-2G(1/r;X^\pm)\,.
    \end{align}
\item For the remaining fluctuations within $S^5$, 
%$\{Z_3,\bar Z_3\}\subset\mathcal{I}$
    \begin{align}
        \delta_{Z_3}(r;X^\pm)\equiv\delta_{\tilde 1\tilde
3}(r;X^\pm)=G(r;X^\pm)-G(1/r;X^\pm),\\  
\delta_{\bar Z_3}(r;X^\pm)\equiv\delta_{\tilde 2\tilde
4}(r;X^\pm)=G(r,X^\pm)-G(1/r;X^\pm)\,.\label{fgbZ3}
    \end{align}
\item For $AdS_5$, 
    \begin{equation}
        \delta_{\hat 1\hat 3}=
        \delta_{\hat 1\hat 4}=
        \delta_{\hat 2\hat 3}=
        \delta_{\hat 2\hat 4}=0\quad\Leftrightarrow\quad
\delta_{I}(r;X^\pm)=0,
    \end{equation} 
for $I\in\mathcal{I}_{AdS_5}\equiv \{Y_2,\bar Y_2, Y_3,\bar Y_3\}$
\item Finally for the eight fermions
$\mathcal{I}=\mathcal{I}_\theta\cup\mathcal{I}_\eta$,
    \begin{equation}
	  \begin{array}{c}
        \delta_{\hat 1\tilde 3}(r;X^\pm)=
        \delta_{\tilde 1\hat 4}(r;X^\pm)=
        \delta_{\hat 2\tilde 3}(r;X^\pm)=
        \delta_{\tilde 2\hat 4}(r;X^\pm)=G(r;X^\pm) \\
			\Updownarrow\\
			\delta_{I}(r;X^\pm)=G(r;X^\pm),\quad
I\in\mathcal{I}_\theta\equiv\{\theta_i\}_{i=1,\dots,4},
	\end{array}\\
    \end{equation}
    \begin{equation}
	\begin{array}{c}
        \delta_{\tilde 1\hat 3}(r;X^\pm)=
        \delta_{\hat 1\tilde 4}(r;X^\pm)=
        \delta_{\tilde 2\hat 3}(r;X^\pm)=
        \delta_{\hat 2\tilde 4}(r;X^\pm)=-G(1/r;X^\pm)\\
		\Updownarrow \\
			\delta_{I}(r;X^\pm)=-G(1/r;X^\pm),\quad
I\in\mathcal{I}_\eta\equiv\{\eta_i\}_{i=1,\dots,4}.
	\end{array}
    \end{equation}
\end{itemize}
In the GM limit $X^{\pm}\rightarrow x^\pm\equiv\exp(\pm ip/2)$ these simplify to
\begin{equation}
\begin{cases}
	\delta_I(r;x^\pm)=0, &I\in\mathcal{I}_{AdS_5},\\
	\delta_I(r;x^\pm)=-2i\log\Bigl(\frac{r-\xp}{r-\xm}\Bigr),
&I\in\mathcal{I}_{S^5},\\
	\delta_I(r;x^\pm)=-i\log\Bigl(\frac{r-\xp}{r-\xm}\Bigr),
&I\in\mathcal{I}_\text{fermions}. 
\end{cases}
\end{equation}
The results obtained in this section thus agree (up to a constant in
the DGM momentum $p$) with the results from the dressing method for the $S^5$
sector - compare (\ref{fgZ2}-\ref{fgbZ3}) with (\ref{dZ2}-\ref{dmZ3})
using \eqref{resolventDGM}. In the GM limit we reproduce also the phase shifts 
determined for the fermions from their explicit solution 
(see Eqn (\ref{dmFerm})). 

\subsection{Phase Shift from $\fsu(2|2)$ S-Matrix}\label{BeisertMatrix}
\paragraph{}
In this subsection, we shall consider yet another way of deriving the 
classical phase shifts for the worldsheet fields in the Giant Magnon
background. We will exploit a relation between the phase shifts and a
particular weak-coupling limit of the exact magnon S-matrix. In
particular, we will take the exact S-matrix for two magnons and take
the Giant Magnon limit for one of the incoming particles and the
plane-wave limit for the other. In this case the first magnon will
become a semiclassical worldsheet soliton and the second an elementary
quantum corresponding to a small fluctuation of the worldsheet fields
around the soliton background. In such a limit the phase of
the S-matrix goes over to the classical phase shift we seek. 
By varying the polarisations of the second magnon we can select 
the phase shift corresponding to each worldsheet field.  Of course 
our ultimate goal is to {\em test} the exact S-matrix at one-loop order so
this may sound like a circular argument. However, the calculation of
the classical phase shift discussed in this subsection 
relies only on the well-tested {\em
  tree-level} contribution to the exact S-matrix (the AFS phase) as
well as the index structure of the S-matrix which is completely
determined by supersymmetry \cite{BeisertSU22}. The calculation
here should be considered as a consistency check for the results
obtained from the finite-gap solution and the dressing method. 
One drawback is that we only know the full S-matrix for ordinary
magnons and not for their bound states\footnote{To do this, 
it will be necessary to apply the fusion procedure to the entire 
$\fpsu(2|2)^{2}\ltimes {\mathbb{R}}^{3}$ magnon scattering matrix, following \cite{CDO2}.}. This means that we can only
extract the phase shifts for scattering in the background of a
charge-less Giant Magnon and not in the more general case of the Dyonic
Giant Magnon described above. On the other hand 
this approach does not require one to choose the
polarisation of the background or the ``large magnon'', as it was
the case in the previous sections, hence the universality of the
semiclassical correction $\theta_{1}(x^{\pm},y^{\pm})$ is more apparent.
\paragraph{}
To begin with let us recall the schematic form for the full scattering
matrix for the elementary magnons given of all sixteen possible
flavors given in \cite{BeisertSU22} 
\begin{equation}
\is(x,y)=s^{0}(x,y)\left[\his(x,y)\otimes\his'(x,y)\right]\,,\label{fullSmatrix}
\end{equation}
where the abelian factor $s^{0}(x,y)$ is given by
\begin{equation}
s^{0}(x,y)=\frac{\xm-\yp}{\xp-\ym}\frac{1-1/\xp\ym}{1-1/\xm\yp}
\sigma^{2}(x,y)\label{Defs0}\,.
\end{equation}
The scattering matrix (\ref{fullSmatrix}) was obtained by demanding
its invariance under the residual symmetry algebra $\fpsu(2|2)\times
\fpsu(2|2)\ltimes {\mathbb{R}}^{3}$, and it has been shown to satisfy
both unitarity and Yang-Baxter equation. To recover the $\fsu(2)$
magnon scattering matrix in (\ref{fullsu2Smatrix}), one simply has to
fix the polarisation of the magnon and isolate the relevant component.
Moreover as argued in \cite{BeisertSU22}, instead of dealing with
all ($16^{4}$) components of (\ref{fullSmatrix}), we can treat the two
copies of $\fpsu(2|2)\ltimes {\mathbb{R}}^{3}$ independently and only
identify their central charges. This greatly reduces the number of the
components we need to deal with to $4^{4}=256$ and we only need to
consider the $\fsu(2|2)$ dynamics scattering matrix $\his(x,y)$.
\paragraph{} 
Recall that the action of $\fsu(2|2)$ dynamic S-matrix
$\his(x_{j},x_{k})$ on a two excitation state is schematically given
by
\begin{equation}
	\his(x_{j},x_{k})|\dots\cX_{j}\cX_{k}'\dots\rangle\to ({\rm{Coeff.}})|
	\dots\cX_{k}''\cX_{j}'''\dots\rangle\,.\label{defShat}
\end{equation}
Here an excitation $\cX_{j}$ with spectral parameters $x^{\pm}_{j}$
can be any component of the $2+2$ dimensional fundamental
representation $\{\p^{1},\p^{2}|\psi^{1},\psi^{2}\}$ of
$\fpsu(2|2)\ltimes \mbR^{3}$.  Notice that in (\ref{defShat}), under
the action of $\his(x_{j},x_{k})$, the momenta/spectral parameters of
the two excitations have been swapped and their flavors are also
allowed to change.  As discussed before, in order to derive the
leading semiclassical correction $\theta_{1}(x,y)$ (\ref{Defk1}) to
the classical dressing phase, we should consider the scattering
between a fluctuation $\bZ$ (or elementary magnon in the plane wave
regime) with spectral parameters $z^{\pm}$ and another arbitrary
elementary magnon $\bX$ with spectral parameters $x^{\pm}$. We can
begin with the full exact expression for the magnon scattering matrix
(\ref{fullSmatrix}) but only keep the lowest order $\theta_{0}(z,x)$
in the dressing phase, which can be readily written as:
\begin{equation}
	\exp(ig\theta_{0}(z,x))=\frac{1-1/\zm\xp}{1-1/\zp\xm}
	\left(\frac{1-1/\zm\xp}{1-1/\zp\xp}
	\frac{1-1/\zp\xm}{1-1/\zm\xm}\right)^{ig(\zeta-u)}\,.
\label{expAFS}
\end{equation}
Here we have introduced the ``rapidity parameters'' $\zeta$ and $u$
\begin{equation}
	\zeta=z+\frac{1}{z}\,,~~~u=x+\frac{1}{x}\,.\label{DefRapidity}
\end{equation}
If we further impose the plane wave limit (\ref{PPlimit2}) on $z^{\pm}$, $\zp\sim\zm=r$,
the scalar factor $s^{0}(r;x)$ (\ref{Defs0}) is then simplified to
\begin{equation}
	s^{0}(r;x)=\frac{r-\xp}{r-\xm}\frac{r-1/\xp}{r-1/\xm}\label{AppS0}\,.
\end{equation}
The $\fsu(2|2)$ scattering matrix $\his(r;x)$ also simplifies
dramatically in the limit (\ref{PPlimit2}), using the in the notations
in (\ref{ExplicitComponents}), the only non-vanishing components are:
\begin{equation}
	a(r;x)=e(r;x)=\frac{r-\xm}{r-\xp}\sqrt{\frac{\xm}{\xp}}
	\,,~~~c(r;x)=-f(r;x)=-1\,.\label{nvcomponents}
\end{equation}
In fact with appropriate choice of the basis for the incoming
excitations, $\his(r;x)$ can be arranged into diagonal form.
\paragraph{}
Substituting (\ref{AppS0}) and (\ref{nvcomponents}) into the full
expression (\ref{fullSmatrix}), we can the easily obtain the
scattering phase between the fluctuation $\bZ$ of different
polarisations and the arbitrary magnon $\bX$. If $\bZ$ belongs to
one of the four bosonic scalar fluctuations $(\p_{1}\tp_{1},
\p_{1}\tp_{2},\p_{2}\tp_{1},\p_{2}\tp_{2})$ which are identified with
string worldsheet fields 
$\{Z_2, \bar{Z}_2, Z_3, \bar{Z}_3\}$ up to linear combinations (see
for example \cite{KMRZ} for more precise identifications), 
its scattering phase with $\bX$ is given by
\begin{equation}
	\delta(r;x^\pm)=-i\log\left(\frac{r-\xp}{r-\xm}\right)+
	i\log\left(\frac{r-1/\xp}{r-1/\xm}\right)+p\,.\label{ExactBphase}
\end{equation}
\paragraph{}
If $\bZ$ belongs to one of the four derivatives fluctuations
$(\psi_{1}\tpsi_{1},\psi_{1}\tpsi_{2},\psi_{2}\tpsi_{1},\psi_{2}\tpsi_{2})$
which can be identified with $\{Y_2, \bar{Y}_2, Y_3, \bar{Y}_3\}$,
its scattering phase with $\bX$ is given by
\begin{equation}
	\del(r;x^\pm)=i\log\left(\frac{r-\xp}{r-\xm}\right)+i\log
	\left(\frac{r-1/\xp}{r-1/\xm}\right)\label{ExactDphase}\,.
\end{equation}
\paragraph{}
Finally, if $\bZ$ belongs to one of the eight fermionic fluctuations
$(\p_{1}\tpsi_{1}\,\p_{1}\tpsi_{2},\p_{2}\tpsi_{1},\p_{2}\tpsi_{2},
\psi_{1}\tp_{1},\psi_{1}\tp_{2},\psi_{2}\tp_{1},\psi_{2}\tp_{2})$ which
can be identified with $\{\theta_1, \dots,\theta_4; \eta_1,\dots, \eta_4\}$,
its scattering phase with $\bX$ given by
\begin{equation}
	\del(r;x^\pm)=i\log\left(\frac{r-1/\xp}{r-1/\xm}\right)+
	\frac{p}{2}\label{ExactFphase}\,.
\end{equation}
Notice that in deriving (\ref{ExactBphase})-(\ref{ExactFphase}), we
have not specify the polarisation of $\bX$; the point is that one
can sure that because of the diagonal form of the reduced
$\fsu(2|2)$ scattering matrix, the phase shifts derived here are in
fact universal and independent of the polarisation of $\bX$.
\paragraph{}
In general, the expressions (\ref{ExactBphase})-(\ref{ExactFphase}) do
not coincide with the exact semiclassical phase-shifts calculated from
the finite gap solution and the dressing method. This can be explained
by the fact that for example in the string sigma model, the exact
phase shift was obtained from scattering with dyonic giant magnon,
which in turns correspond to the $\fsu(2)$ magnon bound states. Here
the approach using $\fsu(2|2)$ scattering matrix is only strictly
valid for the elementary magnons. To make proper comparison with the
exact results from sigma model, one should apply the similar bootstrap
method used in \cite{CDO2} to the various components here and
construct the bound state scattering matrix.  However we do expect the
results here to match when one consider $\bX$ to be in the giant
magnon regime (\ref{GMlimit2}), the exact expressions for the
semiclassical phase shift (\ref{ExactBphase}), (\ref{ExactFphase}) and
(\ref{ExactDphase}) reduce in such limit to 
\begin{eqnarray}
	\del_{I}(r;x^\pm)&=&-2i\log\left(\frac{r-\xp}{r-\xm}\right)
	+p\,,~~~~I\in{\mathcal{I}}_{\mathrm S^{5}}\,,\label{GMBphase}\\
	\del_{I}(r;x^\pm)&=&-i\log\left(\frac{r-\xm}{r-\xp}\right)
	+i\log\left(\frac{r-\xm}{r-\xp}\right)=0\,,~~~~I\in{\mathcal{I}}_{\mathrm AdS_{5}}\,,\label{GMDphase}\\
	\del_{I}(r;x^\pm)&=&-i\log\left(\frac{r-\xp}{r-\xm}\right)
	+\frac{p}{2}\,,~~~~I\in{\mathcal{I}}_{\rm fermions}\,.\label{GMFphase}
\end{eqnarray}
Respectively, (\ref{GMBphase}), (\ref{GMDphase}) and (\ref{GMFphase})
should compare with the phase shifts experienced by giant magnon due
to the scattering with the fluctuations in $S^{5}$, in $AdS_{5}$ and
the fermionic fluctuations; one clearly observes that the expressions
(\ref{GMBphase})-(\ref{GMFphase}) precisely match with the results
from the finite gap solutions and the dressing method up to
linear-momentum dependent terms.

\section{The Zero Energy Shift and the One-loop Correction to the
Dressing Phase}\label{Results}
\paragraph{}
In this section we collect the scattering phases between magnon and
fluctuation calculated from various approaches, and apply the formulae
(\ref{f2}) and (\ref{f4}).  In the present context these become 
(after changing variables from $k$ to $r$ in the integrals), 
\begin{eqnarray}
	\Delta
	E(X^{\pm}) &=&
\frac{1}{2\pi}\sum_{I\in\mathcal{I}}(-1)^{F_{I}}\int^{+1}_{-1}\,dr\frac{\pa
	\delta_{I}(r;X^{\pm})}{\pa r}\sqrt{k(r)^{2}+1}
	\label{Energyshift}\,,\\
	2\Delta
	\Theta(X^{\pm},Y^{\pm})&=
&\frac{1}{2\pi}\sum_{I\in\mathcal{I}}(-1)^{F_{I}}\int^{+1}_{-1}
	\,dr\frac{\pa
	\del_{I}(r;X^{\pm})}{\pa r}
	\del_{I}(r;Y^{\pm})\,\label{oneloopformula}
\end{eqnarray}
where $k(r)=2r/(r^{2}-1)$ and
where the sums are over all possible polarisations for the
intermediate plane-waves, $\mathcal{I}=\mathcal{I}_{AdS_5}\cup\mathcal{I}_{S^5}\cup
\mathcal{I}_\text{fermions}$. The two dyonic giant magnon are
characterised by the spectral data $X^\pm,\,Y^\pm$. The factor of two on the LHS of 
(\ref{oneloopformula}) is related to the normalisation for the 
dressing phase in (\ref{fullSmatrix}). 
We will now use these formulae 
to demonstrate the vanishing one-loop energy shift for the soliton  
and extract the one loop correction to the dressing phase. 
\paragraph{}
It is simple to demonstrate the vanishing energy-shift using the
phase-shifts $\delta_{I}(r;X^\pm)$ calculated in Section 3. We 
then only need to show the weighted summation over $\delta_{I}(r;X^\pm)$
in (\ref{Energyshift}) vanishes up to constant $r$-independent terms.
To perform the calculation, one first notes that the fluctuations with
a polarisation along $AdS_{5}$ will not suffer a phase shift,
\[\delta_I=0,\quad I\in\mathcal{I}_{AdS_5}.\] The weighted summation over the
phase-shifts for the scattering of the four transverse bosonic
fluctuations in $S^{5}$ and the eight fermionic fluctuations becomes,
\begin{multline}
	\sum_{I\in{\mathcal
I}}(-1)^{F_{I}}\delta_{I}(r;X^{\pm})=\overbrace{2G(r;X^\pm)}^{Z_2}
	\underbrace{-2G(1/r,X^\pm)}_{\bar
Z_2}+\\+\underbrace{2[G(r;X^\pm)
	-G(1/r,X^\pm)]}_{Z_3,\bar
Z_3}-[\overbrace{4G(r;X^\pm)}^{\theta_1,\dots,\theta_4}
\underbrace{-4G(1/r,X^\pm)}_{\eta_1,\dots,\eta_4}]=0\,.\label{VPshift}
\end{multline}
We then automatically obtain from \eqref{Energyshift} 
the predicted vanishing of the one-loop energy correction 
for the magnon and its bound states
\footnote{A related calculation appeared in \cite{Spradlin}. In
  particular, it was noted that the range and frequencies of the 
continuous spectra associated with bosonic and fermionic modes were
  the same. However, as we have emphasized above, to compute the
  one-loop correction to the soliton energy it is also necessary to
  determine the appropriate density of states for each mode. See eg 
\cite{Kaul} for an example where this point is essential.}
\begin{equation}
\Delta E=0.
\end{equation}
\paragraph{}
We now move on to the one-loop correction to the soliton S-matrix. 
We are seeking the
equality $\Delta\Theta(X^{\pm},Y^{\pm})=\Theta_{1}(X^{\pm},Y^{\pm})$, 
where $\Theta_{1}(X^{\pm},Y^{\pm})$
is given as
\begin{equation}
	\Theta_{1}(X^{\pm},Y^{\pm})=
K_{1}(\Xp,\Yp)-K_{1}(\Xp,\Ym)-K_{1}(\Xm,\Yp)
	-K_{1}(\Xm,\Ym)\label{DefTheta1}\,.
\end{equation}
Our strategy here is that, instead of comparing with $\Theta_{1}(X,Y)$
using the expression for $K_{1}(X,Y)$ in (\ref{Defk1},\ref{Defk1b}), we shall
consider the derivatives of $\Theta_{1}(X,Y)$ to avoid the issues of
the branch cuts coming from the logarithms.  Differentiating with
respect to $V=(Y^{+}+Y^{-}+1/Y^{+}+1/Y^{-})/2$ we obtain, 
\begin{equation}
	\frac{\pa\Theta_{1}(X^\pm,Y^\pm)}{\pa V}=\frac{\left(F_{1}(\Xp,\Yp)
	-F_{1}(\Xm,\Yp)\right)}{1-1/(\Yp)^{2}}
	+\frac{\left(F_{1}(\Xm,\Ym)-F_{1}(\Xp,\Ym)\right)}{1-1/(\Ym)^{2}}
	\label{DerHLterm}\,,
\end{equation}
where, 
\begin{equation}
	F_{1}(X,Y)=\frac{\pa K_{1}(X,Y)}{\pa Y}
	=\frac{1}{\pi}\left[\frac{1}{Y-X}-\frac{1}{Y-1/X}\right]\log
	\left(\frac{Y+1}{Y-1}\frac{X-1}{X+1}\right)\,,\label{DefF1}
\end{equation}
and we have used the identities $\frac{\pa Y^{\pm}}{\pa
V}=\frac{1}{1-1/(Y^{\pm})^{2}}$.
\paragraph{}
We shall therefore evaluate the corresponding derivative of our
semiclassical result,  
\begin{equation}
2\frac{\pa \Delta\Theta(X^{\pm},Y^{\pm})}{\pa V}
=\frac{1}{2\pi}\sum_{I\in\cI}(-1)^{F_{I}}\int^{+1}_{-1}\,dr \frac{\pa
  \del_{I}(r;X^\pm)}{\pa r}
\frac{\pa \del_{I}(r;Y^\pm)}{\pa V}\,,\label{DdelTheta}
\end{equation}
using the various scattering phases $\del_{I}(r;X^\pm)$ between the
fluctuations and the magnon polarized in one of the $S^{3}\subset
S^{5}$ calculated in the previous sections. Instead of evaluating
every terms in the weighted summation of (\ref{DdelTheta}), again
the four fluctuations in $AdS_{5}$ give vanishing contributions.
Moreover each of the two bosonic fluctuations parallel to the
$S^{3}$ will give four times of the contribution coming from each of
the eight fermionic fluctuations, therefore these contributions
again cancel after taking account of the multiplicities and weights.
As the result, we only need to consider the contributions coming
from the two bosonic fluctuations transverse to the $S^{3}$, i.e., 
$\del_{Z_3}(r;X^\pm)=\del_{\bar Z_3}(r;X^\pm)=G(r;X^\pm)-G(1/r;X^\pm)$.
The relevant derivatives are given by:
\begin{eqnarray}
	&&\frac{\pa\del_{Z_3}(r;X^\pm) }{\pa r}=i\left[\left(\frac{1}{r-\Xp}-
	\frac{1}{r-1/\Xp}\right)-\left(\frac{1}{r-\Xm}-\frac{1}{r-1/\Xm}\right)\right]\,,\label{Drdel}\\
	&&\frac{\pa\del_{Z_3}(r;Y^\pm) }{\pa V}=i\left[\frac{1}{1-1/(\Yp)^{2}}
	\left(\frac{1}{\Yp-r}-\frac{1}{\Yp-1/r}\right)
	-\frac{1}{1-1/(\Ym)^{2}}\left(\frac{1}{\Ym-r}-\frac{1}{\Ym-1/r}\right)\right]\,.\nn\\
\label{Dydel}
\end{eqnarray}
Substituting (\ref{Drdel}) and (\ref{Dydel}) into (\ref{DdelTheta}),
it should be clear that it can be rearranged into
\begin{equation}
	2\frac{\pa \Delta\Theta(X^\pm,Y^\pm)}{\pa V}=2\left[\frac{\tF(\Xp,\Yp)-
	\tF(\Xm,\Yp)}{1-1/(\Yp)^{2}}+\frac{\tF(\Xm,\Ym)-\tF(\Xp,\Ym)}{1
	-1/(\Ym)^{2}}\right]\,,\label{DdelTheta2}
\end{equation}
where the function $\tF(X,Y)$ is given by
\begin{eqnarray}
	\tF(X,Y)&=&\frac{1}{2\pi}\int^{+1}_{-1}\, dr\left[\frac{1}{r-X}-
	\frac{1}{r-1/X}\right]\left[\frac{1}{Y-r}-\frac{1}{Y-1/r}\right]\nn\\
	&=&\frac{1}{\pi}\left[\frac{1}{Y-X}-\frac{1}{Y-1/X}\right]\log
	\left(\frac{Y+1}{Y-1}\frac{X+1}{X-1}\right)
\label{DefTildeF}\,.
\end{eqnarray}
In the second line of (\ref{DefTildeF}) we have used the integrals
(\ref{Aint1}) and (\ref{Aint2}) in the appendix \ref{Integrals}, and
we obtain the exact match between $\tF(X,Y)$ and $F_{1}(X,Y)$ in
(\ref{DerHLterm})!
\paragraph{}
The authors would like
to thank B. Vicedo and L. I. Uruchurtu for usefull discussions, 
they are also grateful to K. Okamura and M. Spradlin for the comments on
the draft.
ND is supported by a PPARC Senior Research Fellowship. RFLM is
supported by the Funda\c c\~ao para a Ci\^encia e Tecnologia with the
fellowship SFRH/BD/16030/2004.

\appendix
\section{Derivation for the one-loop energy-shift
formula}\label{Deriv1}
\paragraph{}
Here we present a derivation for the semiclassical one-loop energy
shift formula (\ref{oneloopformula}). Let us consider a real scalar
field $\vp(x,t)$ in a 1+1 dimensional field theory which contains a
mass parameter $m$ and coupling $g$, we shall consider the strong
coupling limit $g\gg 1$ hence the natural expansion parameter is the
inverse coupling $1/g$. Now suppose the theory admits a classical
one soliton solution $\vp(x,t)\equiv\vp_{cl}(x,t; p)$ where $p$ is
the conserved momentum carried by the soliton, such solution should
have the asymptotic behaviour:
\begin{equation}
\vp_{cl}(x,t; p)\sim \exp(-c|x|)\,,~~~|x|\to\infty\,,\label{asymp1}
\end{equation}
where $c\equiv c(p)$ is the mass of the static soliton at rest. 
The energy of the soliton $E(p,g)$ should also admit the strong
coupling expansion in $1/g$ as
\begin{equation}
E(p)=gE_{cl}(p)+\Delta E(p)+\cO(1/g)\label{Eexpansion}\,,
\end{equation}
where $E_{cl}(p)$ is the classical energy, whereas $\Delta E(p)$ is
the semiclassical one-loop energy shift due to the small quantum
fluctuations around the classical soliton background.
\paragraph{}
To determine $\Delta E(p)$, we first consider the standard small
fluctuation operator in the soliton theory given by
\begin{equation}
	\hat{H}=\frac{\delta^{2}\cL(\vp,\pa\vp)}{\delta\vp^{2}(x,t)}
	\vline_{\vp=\vp_{cl}(x,t; p)}\label{DefH}\,,
\end{equation}
where $\cL(\vp,\pa\vp)$ is the Lagrangian of the theory. The
semiclassical energy shift $\Delta E(p)$ is then determined by the
spectrum of $\hat{H}$; asymptotically, i.e. away from the soliton,
$\hat{H}$ should tend to quantum mechanical Hamiltonian describing the
propagation of plane wave: 
\begin{equation}
	\hat{H}\to \Box+m^{2}+\cO\left(e^{-c|x|}\right)\,,~~~|x| \to 
	\infty\,,\label{asym2}
\end{equation}
where $\Box=-\pa_t^2+\pa_x^2$.
Hence if we consider a solution $\psi(x,t)$ to the linearised equation
of motion, i.e. it satisfies
\begin{equation}
	\hat{H}\psi(x,t;k)=0\,,~~~\psi(x,t;k)\in{\mathbb{C}}\,.
\end{equation}
Asymptotically, to be consistent with (\ref{asym2}), the solution
$\psi(x,t)$ should have the following behaviour:
\begin{eqnarray}
	&&\psi(x,t;k)\to \exp(iE(k)t+ikx)\,,~~~x\to -\infty\,,\nn\\
	&&\psi(x,t;k)\to \exp(i\delta(k;p)+iE(k)t+ikx)\,,~~~x\to\infty\,,
\label{asym3}
\end{eqnarray}
where $k$ is the wave vector of $\psi(x,t;k)$ and $\epsilon(k)$ is
an eigenvalue of the asymptotic Hamiltonian (\ref{asym2}), so that
$E(k)=\sqrt{k^{2}+m^{2}}$. As it propagates from $x=-\infty$
to $x=\infty$, the fluctuation $\psi(x,t;k)$ scatters elastically
with the classical soliton $\vp_{cl}(x,t;p)$, the unitarity of
$\hat{H}$ demands that such scattering can only introduce an overall
phase-shift $\delta(k;p)$ into $\psi(x,t;k)$,
$\delta(k;p)$ is called the ``scattering phase''\footnote{In our
analysis, we exclude the possible formation of bound states, and we
assume that there is no reflection, however they are indeed true in
the case of our interests.}.
\paragraph{}
We now would like to derive the one-loop energy shift $\Delta E$ of
the soliton due to the presence of the fluctuation $\psi(x,t;k)$.
Instead of considering an infinite line, we now impose periodic
boundary condition on the soliton wave function $\vp_{cl}(x,t;p)$,
i.e.
\begin{equation}
	\vp_{cl}(x,t;p)=\vp_{cl}(x+L,t;p)\,,~~~L\gg 1\label{PBC1}\,;
\end{equation}
as the result the fluctuation $\psi(x,t;k)$ also acquires the periodicity:
\begin{equation}
\psi(x,t;k)=\psi(x+L,t;k)\,.\label{PBC2}
\end{equation}
Comparing (\ref{PBC2}) with the asymptotic condition earlier
(\ref{asym3}), we can deduce that the allowed valued of wave vector
$k_{n}$ must satisfy the condition
\begin{equation}
	k_{n}L=2\pi n+\delta(k_{n};p)\,,~~~n\in{\mathbb{Z}}\label{PBC3}\,.
\end{equation}
Typically we expect that for a given wave vector $k=k_{n}$, there
should be an unique solution.  We can actually impose similar periodic
boundary condition in the time direction on the soliton, that is for
some given time period $T$,
\begin{equation}
	\vp_{cl}(x,t;p)=\vp_{cl}(x,t+T;p)\,.\label{PBC4}
\end{equation}
Whereas for the fluctuation $\psi(x,t;k)$, after one period $T$, it
picks up a phase given by
\begin{equation}
\psi(x,t+T;k)=\exp(i\nu(k))\psi(x,t;k)\label{PBC5}\,,
\end{equation}
where $\nu(k)=E(k)T=\sqrt{k^{2}+m^{2}}T$, the phase $\nu(k)$ is
called ``stability angle'' in the literature.
\paragraph{}
Essentially, the derivation for the one-loop energy-shift boils down
to comparing the stability angles in the vacuum (without the
presence of soliton) and with the existence of soliton. In the
vacuum, we can write down the stability angle:
\begin{eqnarray}
\nu(k_{n}^{(0)})&=&=E(k^{(0)}_{n})T=\sqrt{\left(k_{n}^{(0)}\right)^{2}
	+m^{2}}T\,,\label{SA1}\\
	Lk_{n}^{(0)}&=&2\pi n\,,~~~n\in{\mathbb{Z}}\,.\label{PBC6}
\end{eqnarray}
Here $k^{(0)}_{n}$ denotes the wave vector for the plane wave
propagating in the vacuum and the equation (\ref{PBC6}) is simply
the consequence of the periodicity in $x$-direction. In the soliton
background, we can again write down the stability angle for the
fluctuation:
\begin{equation}
\nu(k_{n})=\sqrt{k_{n}^{2}+m^{2}}T\,,\label{SA2}
\end{equation}
with the wave vector $k_{n}$ now satisfies the periodic condition
(\ref{PBC3}).  In \cite{Dashen}, the general formula for the one loop
energy shift such time-dependent solution is given simply as
\begin{eqnarray}
	\Delta
	E_{L}(p)&=&\sum^{+\infty}_{n=-\infty}\left(\frac{\pa\nu(k,T)}{\pa
	T}\vline_{k=k_{n}}-\frac{\pa\nu(k,T)}{\pa
	T}\vline_{k=k_{n}^{(0)}}\right)\nn\\
	&=&\sum_{n=-\infty}^{+\infty}\left(\sqrt{k_{n}^{2}+m^{2}}-
	\sqrt{\left(k_{n}^{(0)}\right)^{2}+m^{2}}\right)\label{Eshift1}
\end{eqnarray}
As we take the continuous $L\to\infty$ limit, $k_{n}=\frac{2\pi
n}{L}+\cO(1/L)$ for high mode numbers $|n|\sim L$, simple algebra
shows that $E(k_{n})=E(k_{n}^{(0)})+\cO(1/L)$. In such
limit, the summation over the mode number $n$ goes over to an
integral, however we can also equivalently express it as integral
over the wave vector $k$, to do so we need to write down the density
of states in the soliton background and in the vacuum defined to be:
\begin{equation}
\frac{\pa n}{\pa k}=\frac{L}{2\pi}+\frac{1}{2\pi}\frac{\pa \delta(k;p)}{\pa k}\,,~~~
\frac{\pa n}{\pa k^{(0)}}=\frac{L}{2\pi}\,.\label{DOS}
\end{equation}
Finally we deduce the one-loop energy shift formula (\ref{Eshift1})
goes over to
\begin{eqnarray}
	\Delta E(p)&=&\lim_{L\to\infty}\left[\Delta E_{L}(p)\right]
	=\int^{+\infty}_{-\infty}\,dk\left(\frac{\pa n}{\pa k}-
	\frac{\pa n}{\pa k^{(0)}}\right)\sqrt{k^{2}+m^{2}}\nn\\
	&=&\frac{1}{2\pi}\int^{+\infty}_{-\infty}\,dk\frac{\pa \delta(k;p)}{\pa k}
	\sqrt{k^{2}+m^{2}}\,.\label{Eshift2}
\end{eqnarray}
For the case of $N_F$ decoupled real fluctuation fields
$\psi_{I}(x,t;k)\,~I=1,\dots,N_F$ (include bosonic and fermionic
fields), the generalisation is obvious. Furthermore if they all share
the same dispersion relations as it is true for the plane wave magnon
we consider in this paper, the formula gets extra simplifications,
taking into the account of opposite weighting for the bosons and
fermions, we finally derive the one loop energy shift formula
(\ref{oneloopformula}): 
\begin{equation}
	\Delta E(p)=\frac{1}{2\pi}\sum_{I=1}^{N_F}(-1)^{F_{I}}\int^{+\infty}_{-\infty}
	\,dk\frac{\pa \delta_{I}(k;p)}{\pa k}\sqrt{k^{2}+m^{2}}\,,\label{Eshift3}
\end{equation}
where $\delta_{I}(k;p)$ corresponds to the scattering phase between
the $I$-th fluctuation and the soliton.

\section{Derivation for the one-loop phase shift formula}\label{Deriv2}
\paragraph{}
In this appendix we present the derivation for the formulae of
one-loop corrections to the scattering phase given in the equations
(\ref{f3}) and (\ref{f4}).  As in the main text, we begin by
considering a two soliton solution with momenta $p_{1}$ and $p_{2}$
respectively in a $1+1$ dimensional field theory characterised by
coupling constant $g$, this configuration can be described by a
scattering wave function $\varphi_{scat}(x,t; x_1^{(0)},x_2^{(0)},p_{1},p_{2})$. In
addition we also impose the periodic boundary condition:
\begin{equation}
	x\sim x+ L\,,~~~~\varphi_{scat}(x,t; x_1^{(0)},x_2^{(0)},p_{1},p_{2})\sim 
	\varphi_{scat}(x+L,t; x_1^{(0)},x_2^{(0)},p_{1},p_{2})~~~~L\gg 1\,,\label{PDC1}
\end{equation}
this also implies the energy levels of the two solitons are quantised.
As the scattering between the two solitons is elastic, the total
energy of the system is given by
\begin{equation}
	E(n_{1},n_{2})\equiv E(p_{n_{1}},p_{n_{2}})=E(p_{n_1})+E(p_{n_2})\,,~~~~n_{1},n_{2}\in 
	{\mathbb{Z}}\label{TotalE}\,,
\end{equation} 
Here $n_{1}$ and $n_{2}$ are again the mode numbers of the two
solitons, $E(p_{n_1})$ and $E(p_{n_2})$ their energies, whereas the
quantised soliton momenta $p_{n_{1}}$ and $p_{n_{2}}$ are given by
\begin{eqnarray}
p_{n_{1}}L&=&2\pi n_{1}+\Theta(p_{n_{1}},p_{n_{2}})\,,\label{PDC2}\\
p_{n_{2}}L&=&2\pi n_{2}-\Theta(p_{n_{1}},p_{n_{2}})\,.\label{PDC3}
\end{eqnarray}
The function $\Theta(p_{n_{1}},p_{n_{2}})$ is the scattering phase
between the two solitons, which in general has strong expansion in
$1/g$ as given in (\ref{expTheta}), and our aim here is to derive a
formula for $\Delta\Theta(p_{n_{1}},p_{n_{2}})$. Notice that the
system also has another natural expansion parameter, namely $1/L$ with
$L\gg 1$; essentially the set-up of our derivation for
$\Delta\Theta(p_{n_{1}},p_{n_{2}})$ is to consider the appropriate
double expansions in both $1/g$ and $1/L$ for the soliton momenta and
energies $p_{n_{i}}$ and $E(p_{n_{i}})\,,~i=1,2$, and apply
(\ref{PDC2}) and (\ref{PDC3}) to relate and identify the terms
associated with $\Delta\Theta(p_{n_{1}},p_{n_{2}})$.
\paragraph{}
Let us begin by expanding the two soliton momenta in $1/L$ while
keeping $g$ fixed, we can then write down:
\begin{equation}
	p_{n_{i}}=p_{n_{i}}^{(0)}+\frac{1}{L}p_{n_{i}}^{(1)}+\cO(1/L^{2})
	\,,~~~i=1,2\,.\label{expPL}
\end{equation}
If we also divide both sides of (\ref{PDC2}) and (\ref{PDC3}) and
replace the momenta entering $\Theta(p_{n_{2}},p_{n_{2}})$ with
(\ref{expPL}), we can obtain that
\begin{equation}
	p_{n_{1}}^{(0)}=\frac{2\pi n_{1}}{L}\sim \cO(1) \,,~~~~p_{n_{2}}^{(0)}
	=\frac{2\pi n_{2}}{L}\sim \cO(1)\,,\label{P10P20} 
\end{equation}
at the leading order and here we have assumed that the mode numbers
$n_{i}$ to be large so that $n_{i}/L$ is kept fixed;  at the next
leading order in $1/L$ expansion we identify that
\begin{equation}
	p_{n_{1}}^{(1)}=-p_{n_{2}}^{(1)}
	=\Theta(p_{n_{1}}^{(0)},p_{n_{2}}^{(0)})=
	g\Theta_{cl}(p_{n_{1}}^{(0)},p_{n_{2}}^{(0)})+
	\Delta\Theta(p_{n_{1}}^{(0)},p_{n_{2}}^{(0)})+\cO(1/g)\,,\label{P11}\\
\end{equation}
We can also perform a similar expansion for the total energy of the
system, which we shall write it as:
\begin{equation}
	E(n_{1},n_{2})=E^{(0)}(n_{1},n_{2})+
	\frac{1}{L}E^{(1)}(n_{1},n_{2})+\cO(1/L^{2})\label{expE}\,,
\end{equation}
again using (\ref{expPL}) we can write down 
\begin{eqnarray}
	E^{(0)}(n_{1},n_{2})&=&E(p_{n_{1}}^{(0)})+E(p_{n_{2}}^{(0)})\label{E0}\,,\\
	E^{(1)}(n_{1},n_{2})&=&g\frac{\pa E_{cl}(p_{n_{1}})}{\pa p_{n_{1}}}
	\vline_{p_{n_{1}}=p_{n_{1}}^{(0)}}\times p_{n_{1}}^{(1)}+
	\frac{\pa E_{cl}(p_{n_{2}})}{\pa p_{n_{2}}}\vline_{p_{n_{2}}=
	p_{n_{2}}^{(0)}}\times p_{n_{2}}^{(1)}\label{E1}
\end{eqnarray}
\paragraph{}
Having expanded in the $1/L$ for the energy, we can now perform
further $1/g$ expansions for (\ref{E0}) and (\ref{E1}), which are can
be written as
\begin{eqnarray}
	E^{(0)}(n_{1},n_{2})&=& g E^{(0)}_{cl}(n_{1},n_{2})
	+\Delta E^{(0)}(n_{1},n_{2})+\cO(1/g)\label{E0g}\,,\\
	E^{(1)}(n_{1},n_{2})&=& g E^{(1)}_{cl}(n_{1},n_{2})
	+\Delta E^{(1)}(n_{1},n_{2})+\cO(1/g)\label{E1g}\,.
\end{eqnarray}
Using the similar double expansion for the energy of individual
soliton, we can rewrite the various quantities in (\ref{E0g}) and
(\ref{E1g}) as the following:
\begin{eqnarray}	E^{(0)}_{cl}(n_{1},n_{2})&=&E^{(0)}_{cl}(n_{1})+E^{(0)}_{cl}(n_{2})\,,\label{E00}\\
	\Delta E^{(0)}(n_{1},n_{2})&=&\Delta E^{(0)}(n_{1})+\Delta E^{(0)}(n_{2})\,,\label{DE0}\\
	E^{(1)}_{cl}(n_{1},n_{2})&=&g\left[\frac{\pa E_{cl}(p_{n_{1}})}{\pa p_{n_{1}}}
	\vline_{p_{n_{1}}=p_{n_{1}}^{(0)}}-\frac{\pa E_{cl}(p_{n_{2}})}{\pa p_{n_{2}}}
	\vline_{p_{n_{2}}=p_{n_{2}}^{(0)}}\right]\Theta(p_{n_{1}}^{(0)},p_{n_{2}}^{(0)})\,,\label{E11}\\
	\Delta E^{(1)}(n_{1},n_{2})&=&g\left[\frac{\pa E_{cl}(p_{n_{1}})}{\pa p_{n_{1}}}
	\vline_{p_{n_{1}}=p_{n_{1}}^{(0)}}-\frac{\pa E_{cl}(p_{n_{2}})}{\pa p_{n_{2}}}\vline_{p_{n_{2}}
	=p_{n_{2}}^{(0)}}\right]\Delta\Theta(p_{n_{1}}^{(0)},p_{n_{2}}^{(0)})\,.\label{DE1}
\end{eqnarray} 
In (\ref{DE1}) we have used the relation (\ref{P11}) and from (\ref{DE1})
we conclude that we can in fact extract the one loop correction to the
soliton scattering phase
$\Delta\Theta(p_{n_{1}}^{(0)},p_{n_{2}}^{(0)})$ from the $1/L$
expansion of the one loop energy $\Delta E(n_{1},n_{2})$!  This useful
observation allows us to recycle the idea used in deriving the
one-loop energy shift for single soliton, that is to consider a plane
wave fluctuation with wave vector $k_{n}$ in the background of two
solitons, and we can denote the total one-loop energy shift to be:
\begin{equation}
	\Delta E(n_{1},n_{2})=\sum_{n=-\infty}^{+\infty}
	\Delta E_{n}(n_{1},n_{2})\,,~~~~n\in {\mathbb{Z}}\,,\label{DelEE}
\end{equation}
where $n$ is the mode number for the plane wave fluctuation.  We
assume the plane wave again scatters elastically with the two
solitons, moreover the classical integrability of the system persists
here, so that the three body scattering matrix can be factorised into
pair-wise scatterings. We can therefore, at the classical level, write
down the periodicity condition for the new three body system:
\begin{eqnarray}
	k_{n}L&=&2\pi n+\delta(k_{n},p_{n_{1}})+
	\delta(k_{n},p_{n_{2}})\,,\label{3kn}\\
	p_{n_{1}}L&=&2\pi n_{1}+g\Theta_{cl}(p_{n_{1}},p_{n_{2}})
	-\delta(k_{n},p_{n_{1}})\,,\label{3P1}\\
	p_{n_{2}}L&=&2\pi n_{2}-g\Theta_{cl}(p_{n_{1}},p_{n_{2}})
	-\delta(k_{n},p_{n_{1}})\,,\label{3P2}
\end{eqnarray}
where $\delta(k_n,p_{n_1})$ and $\delta(k_n,p_{n_2})$ are the
scattering phases between the plane wave and the first and second
soliton respectively.  The $1/L$ expansion in this system yields the
expression for $k_{n}$
\begin{equation}
k_{n}=k_{n}^{(0)}+\frac{1}{L}k_{n}^{(1)}+\cO(1/L^{2})\label{expLk}\,,
\end{equation}
and we can use the similar arguments for obtaining (\ref{P10P20}) and
(\ref{P11}) to deduce in this three body case:
\begin{eqnarray}
	k_n^{(0)}&=&\frac{2\pi n}{L}\,,~~~~p_{n_1}^{(0)}=
	\frac{2\pi n_1}{L}\,,~~~~p_{n_2}^{(0)}=\frac{2\pi n_2}{L}\,,\label{kP1P2}\\
	k_{n}^{(1)}&=&\delta(k_{n}^{(0)},p_{n_{1}}^{(0)})
	+\delta(k_{n}^{(0)},p_{n_{2}}^{(0)})\,,\label{3kn1}\\
	p_{n_1}^{(1)}&=&g\Theta_{cl}(p_{n_{1}},p_{n_{2}})-
	\delta(k_{n}^{(0)},p_{n_{1}})\,,\label{3Pn1}\\
	p_{n_2}^{(1)}&=&-g\Theta_{cl}(p_{n_{1}},p_{n_{2}})-
	\delta(k_{n}^{(0)},p_{n_{2}})\,.\label{3Pn2}
\end{eqnarray}
Here in writing out $p_{n_1}$ and $p_{n_2}$ we have not used
$p_{n_1}^{(0)}$ and $p_{n_2}^{(0)}$, the point is that we will
eventually take the $L\to \infty$ limit, the distinction between them
vanish. However for $k_{n}$ and $k_{n}^{(0)}$, as we will sum over all
infinite mode numbers $-\infty<n<+\infty$ and we expect the summation
to go over the integral in the continuous limit, we should therefore
be careful with the difference even in such limit.  Finally using
above, we can write down the $1/L$ expansion for the total energy
$E_{n}(n_1,n_2)$ of this three body system as
\begin{eqnarray}
&& E_n(n_1,n_2)=\left[g E_{cl}(p_{n_1}^{(0)})+g
E_{cl}(p_{n_2}^{(0)})+\sqrt{(k_{n}^{(0)})^{2}+m^{2}}\right]\nn\\
&&+\frac{g}{L}\left[\frac{\pa E_{cl}(p_{n_{1}})}{\pa
p_{n_{1}}}\vline_{p_{n_{1}}=p_{n_{1}}^{(0)}}\times
p_{n_1}^{(1)}+\frac{\pa E_{cl}(p_{n_{2}})}{\pa
p_{n_{2}}}\vline_{p_{n_{2}}=p_{n_{2}}^{(0)}}\times
p_{n_2}^{(1)}\right]\nn\\
&&+\sqrt{k_{n}^{2}+m^{2}}-\sqrt{(k_{n}^{(0)})^{2}+m^{2}}\nn\\
&&-\frac{g}{L}\left[\frac{\pa E_{cl}(p_{n_{1}})}{\pa
p_{n_{1}}}\vline_{p_{n_{1}}=p_{n_{1}}^{(0)}}
\left(\delta(k_n,p_{n_1})-\delta(k_n^{(0)},p_{n_1})\right)+\frac{\pa
E_{cl}(p_{n_{2}})}{\pa
p_{n_{2}}}\vline_{p_{n_{2}}=p_{n_{2}}^{(0)}}\left(\delta(k_n,p_{n_2})
-\delta(k_n^{(0)},p_{n_2})\right)\right]\,.\label{longexp}\nn\\
\end{eqnarray}
The one-loop energy shift for the two solitons due to the plane wave
of wave vector $k_{n}$ are contained within the last two lines of
(\ref{longexp}), summing over all mode numbers, the total one-loop
energy shift due the plane wave is then given by
\begin{eqnarray}
&&\Delta
E(n_1,n_2)=\sum^{+\infty}_{n=-\infty}\left[\sqrt{k_{n}^{2}+m^{2}}
-\sqrt{(k_{n}^{(0)})^{2}+m^{2}}\right]\nn\\
&&-\frac{g}{L}\sum^{+\infty}_{n=-\infty}\left[\frac{\pa
E_{cl}(p_{n_{1}})}{\pa
p_{n_{1}}}\vline_{p_{n_{1}}=p_{n_{1}}^{(0)}}\left(\delta(k_n,p_{n_1})
-\delta(k_n^{(0)},p_{n_1})\right)+\frac{\pa
E_{cl}(p_{n_{2}})}{\pa
p_{n_{2}}}\vline_{p_{n_{2}}=p_{n_{2}}^{(0)}}\left(\delta(k_n,p_{n_2})
-\delta(k_n^{(0)},p_{n_2})\right)\right]\,.\label{DelEE1}\nn\\
\end{eqnarray}
If we compare (\ref{DelEE1}) with the $1/L$ expansion of $\Delta
E(n_1,n_2)$ (\ref{DelEE}):
\begin{equation}
\Delta E(n_1,n_2)=\Delta E^{(0)}(n_1,n_2)+
\frac{1}{L}\Delta E^{(1)}(n_1,n_2)+\cO(1/L^{2})\label{expDelEE}\,,
\end{equation}
as well as apply the explicit expressions (\ref{DE0}) and (\ref{DE1}),
we can therefore deduce that in the $L\to\infty$ limit
\begin{eqnarray}
\Delta E(n_1,n_2)&=&\lim_{L\to\infty}\sum_{n=-\infty}^{+\infty}
\left[\sqrt{k_{n}^{2}+m^{2}}-\sqrt{(k_{n}^{(0)})^{2}+m^{2}}\right]\,,\label{Eshift0}\\
\Delta\Theta(p_{1},p_{2})&=&\lim_{L\to\infty}\sum_{n=-\infty}^{+\infty}
\left[-\delta(k_n,p_{n_1})+\delta(k_n^{(0)},p_{n_1})\right]\,,\label{Phaseshift1}\\
\Delta\Theta(p_{1},p_{2})&=&\lim_{L\to\infty}\sum_{n=-\infty}^{+\infty}
\left[\delta(k_n,p_{n_2})-\delta(k_n^{(0)},p_{n_2})\right]\,.\label{Phaseshift2}
\end{eqnarray}
The second and third lines above can be calculated independently and
used as a consistency check. 
\paragraph{}
To obtain the integral expressions for
(\ref{Eshift0})-(\ref{Phaseshift2}), we can recycle the arguments in
section \ref{Deriv1} and write down the density of states in this case
\begin{equation}
\frac{\pa n}{\pa k}=\frac{L}{2\pi}+\frac{1}{2\pi}\frac{\pa \delta(k,p_1)}{\pa k }
+\frac{1}{2\pi}\frac{\pa \delta(k,p_2)}{\pa k }\,,~~~~\frac{\pa n}{\pa k^{(0)}}=\frac{L}{2\pi}\,.\label{DOS2}
\end{equation}
Finally plugging in the expressions in (\ref{DOS2}), we can rewrite
(\ref{Eshift0}) into 
\begin{eqnarray}
\Delta E(n_1,n_2)&=&\lim_{L\to\infty}\sum_{n=-\infty}^{+\infty}
\left[\sqrt{k_{n}^{2}+m^{2}}-\sqrt{(k_{n}^{(0)})^{2}+m^{2}}\right]\nn\\
&=&\frac{1}{2\pi}\int^{+\infty}_{-\infty}dk\left(\frac{\pa \delta(k,p_1)}{\pa k }
+\frac{\pa \delta(k,p_2)}{\pa k }\right)\sqrt{k^{2}+m^{2}}\nn\\
&=&\Delta E(p_1)+\Delta E(p_2)\label{Eshift02}\,.
\end{eqnarray}
In this last line of (\ref{Eshift02}) we have used the one-loop energy
shift formula for single soliton we derived earlier (\ref{Eshift2}).
Moreover we can use (\ref{DOS2}) to rewrite
\begin{eqnarray}
\Delta\Theta(p_{1},p_{2})&=&\lim_{L\to\infty}\sum_{n=-\infty}^{+\infty}
\left[-\delta(k_n,p_{n_1})+\delta(k_n^{(0)},p_{n_1})\right]\nn\\
&=&-\frac{1}{2\pi}\int^{+\infty}_{-\infty}dk\left(
\frac{\pa \delta(k,p_1)}{\pa k }+\frac{\pa \delta(k,p_2)}{\pa k }\right)\delta(k, p_1)\nn\\
&=&-\frac{1}{2\pi}\int^{+\infty}_{-\infty}dk\left[\frac{1}{2}
\frac{\pa}{\pa k}\left[\delta^{2}(k,p_1)\right]+\frac{\pa}{\pa k}
\left[\delta(k,p_1)\delta(k,p_2)\right]-\frac{\pa \delta(k,p_1)}{\pa k}\delta(k,p_2)\right]\nn\\
&=&\frac{1}{2\pi}\int^{+\infty}_{-\infty}dk 
\frac{\pa \delta(k,p_1)}{\pa k}\delta(k,p_2)\label{Phaseshift11}
\end{eqnarray}
In the third line of (\ref{Phaseshift11}), we have discarded the total
derivative terms; we can also perform similar calculation for
(\ref{Phaseshift2}) and show that it is identical to
(\ref{Phaseshift11}). In either case, they are indeed the one loop
phase shift for the fluctuation of single flavor given in (\ref{f3}).
For the generalisation, we can consider plane wave fluctuations of
different flavors and both bosonic and fermionic, all of them share
the same same dispersion relations, we can at last write down the
generalised scattering one-loop scattering phase shift:
\begin{equation}
	\Delta\Theta(p_1,p_2)=\frac{1}{2\pi}\sum_{I=1}^{N_F}(-1)^{F_I}\int^{+\infty}_{-\infty}
dk \frac{\pa \delta_{I}(k,p_1)}{\pa k}\delta_{I}(k,p_2)\label{GPhaseshift}\,,
\end{equation}
which was stated and used in the main text (c.f. (\ref{f4})) and this
completes our derivation.

\section{Calculation details for the dressing method}\label{dressingup}

\paragraph{}

In this appendix we present the calculations for the phase shifts
suffered by a plane-wave fluctuation as it scatters a $N$-soliton DGM
string solution lying inside a $S^3$ subspace of the $S^5$ using
dressing method. 
The key equation for deriving the asymptotics of the plane-wave
solution $\delta\g_N$ is given by: 
\begin{equation}\label{deltag}
\delta
\g_N\Big|_{x\rightarrow\pm\infty}=-2i\sin\Bigl(\frac{q}2\Bigr)\Bigl(\mathcal{P}_{N+1}[\tilde
w]+\mathcal{Q}_{N+1}[\tilde
w]\Bigr)\Big|_{\tilde\eta=0,x\rightarrow\pm\infty}\g_N\Bigl|_{x\rightarrow\pm\infty}\,,
\end{equation}
where $\tilde{w}$ is the polarisation vector of the perturbation, $q$ the
perturbation momentum and $\g_N$ the $N$-soliton background solution. We thus have to
determine the asymptotics for both the $N$-soliton solution and for the
projectors
$\mathcal{P}_{N+1}|_{\tilde\eta=0},\mathcal{Q}_{N+1}|_{\tilde\eta=0}$.

\paragraph{}
Taking the asymptotic limit $x\rightarrow\pm\infty$ simplifies the
calculation greatly, since we find that
\begin{equation}\label{projectores1}
    \mathcal{P}_1|_{\pm\infty}\approx\begin{pmatrix} 
	\mathcal{P}_1^{\SU(2)}|_{\pm\infty} &
	0 \\ 0 & 0  \end{pmatrix}, \quad
    \mathcal{Q}_1|_{\pm\infty}\approx\begin{pmatrix} 0 & 0 \\ 0 &
	\mathcal{P}_1^{\SU(2)}|_{\pm\infty} \end{pmatrix},
\end{equation}
where $\mathcal{P}^{\SU(2)}_1|_{+\infty}\approx\begin{pmatrix} 0 & 0 \\ 0 &
1\end{pmatrix}$ and
$\mathcal{P}^{\SU(2)}_1|_{-\infty}\approx\begin{pmatrix} 1 & 0 \\ 0 &
0\end{pmatrix}$ are the asymptotic limits of the projector of the
$\SU(2)$ closed sector and that
\begin{equation}\label{projectores2}
    \mathcal{P}_N|_{\pm\infty}=\mathcal{P}_1|_{\pm\infty},\quad
	\mathcal{Q}_N|_{\pm\infty}=\mathcal{Q}_1|_{\pm\infty},
\end{equation}
when all $N$-solitons have the same polarisation
$w_N=\cdots=w_1=(i,1,0,0)$.
\paragraph{}
The $N$-soliton solution can then be reconstructed from
$\Psi_0$, and written in the following factorised form:
\[\Psi_N( X)=\chi_N( X)\chi_{N-1}( X)\cdots\chi_1( X)\Psi_0( X),\]
with
\[\chi_k(X)=1+\frac{ X_k- \bar{X}_k}{ X- X_k}\mathcal{P}_k[w_k]+
\frac{1/\bar{X}_{1}-1/ X_k}{ X-1/\bar{X}_{k}}\mathcal{Q}_k[w_k].\]
In particular we will have %{\tt (What are $\chi_{k}$??)}
\begin{equation}
    \chi_k( X)|_{+\infty} \approx
	\begin{pmatrix}
	1 & 0 & 0 & 0\\
	0 & \frac{ X- \Xm_k}{ X-\Xp_k} & 0 & 0 \\
	0 & 0 & 1 & 0\\
	0 & 0 & 0 & \frac{ X-1/ \Xp_k}{ X-1/\Xm_k}\end{pmatrix},\qquad
	\chi_k( X)|_{-\infty} \approx
	\begin{pmatrix}
	\frac{ X- \Xm_k}{ X- \Xp_k} & 0 & 0 & 0 \\
	0 & 1 & 0 & 0\\
	0 & 0 & \frac{ X-1/ \Xp_k}{ X-1/\Xm_k} & 0 \\
	0 & 0 & 0 & 1
	\end{pmatrix}
\end{equation}
and so that %{\tt (More explanations here})
\begin{align}\label{psinlima}
	\Psi_N(X)\Big|_{+\infty}&\approx
	\mathrm{diag}\left(e^{iZ(X)}~,\prod_{k=1}^N\frac{ X- \Xm_k}{ X-
	\Xp_k}e^{-iZ(X)}~,e^{iZ(X)}~, \prod_{k=1}^N\frac{ X-1/ \Xp_k}{ X-1/\Xm_k}e^{-iZ(X)}~\right)
	,\qquad
	\\\label{psinlimb}
	\Psi_N(X)\Big|_{-\infty}&\approx
	\mathrm{diag}\left(\prod_{k=1}^{N}\frac{ X- \Xm_k}{X-\Xp_k}e^{iZ(X)}~,e^{-iZ(X)}~,
	\prod_{k=1}^{N}\frac{ X-1/\Xp_k}{ X-1/\Xm_k}e^{iZ(X)}~,e^{-iZ(X)}~\right).
\end{align}
For real $X=\bar{X}=r$ one gets
\begin{align}\label{psinlim1}
\Psi_N(r)\Big|_{+\infty}&\approx
\mathrm{diag}\left(e^{i\frac{v}{2}}~,\prod_{k=1}^N\frac{ r- \Xm_k}{ r-
\Xp_k}e^{-i\frac{v}{2}}~,e^{i\frac{v}{2}}~, 
\prod_{k=1}^N\frac{ r-1/ \Xp_k}{ r-1/\Xm_k}e^{-i\frac{v}{2}}~\right),\qquad
\\\label{psinlim2}
\Psi_N(r)\Big|_{-\infty}&\approx
\mathrm{diag}\left(\prod_{k=1}^{N}\frac{ r- \Xm_k}{r-\Xp_k}e^{i\frac{v}{2}}~,
e^{-i\frac{v}{2}}~,\prod_{k=1}^{N}\frac{ r-1/\Xp_k}{ r-1/\Xm_k}e^{i\frac{v}{2}}~,e^{-i\frac{v}{2}}~\right),
\end{align}
where
\[v\equiv Z(X)+\bar Z(X) = 2Z(r)=\omega t-kx,\] with
$\omega=\sqrt{k^2+1}$ and $k=2r/(1-r^2)$. If we have taken $X=
\bar{X}=1/r$, we would get an identical set of expressions but with $k\rightarrow-k$
(and with $r\rightarrow 1/r$).
\paragraph{}
If one takes $X=0$ in (\ref{psinlima}) and (\ref{psinlimb}), they reduce to
\begin{align}
\g_N|_{+\infty}&=\mathrm{diag}\Bigl(e^{it},e^{-it-iP},e^{it},e^{-it-iP}\Bigr),\\
\g_N|_{-\infty}&=\mathrm{diag}\Bigl(e^{it-iP},e^{-it},e^{it-iP},e^{-it}\Bigr),
\end{align}
where $\sum_{k=1}^N p_k=P$ is the total momentum. We can always
re-scale $\g_N$ by $e^{i\frac{P}2}$ to get a more symmetrical
expression,
\begin{align}
\g_N|_{+\infty}&=\mathrm{diag}\Bigl(e^{it+i\frac{P}2},
e^{-it-i\frac{P}2},e^{it+i\frac{P}2},e^{-it-i\frac{P}2}\Bigr),\\
\g_N|_{-\infty}&=\mathrm{diag}\Bigl(e^{it-i\frac{P}2},
e^{-it+i\frac{P}2},e^{it-i\frac{P}2},e^{-it+i\frac{P}2}\Bigr).
\end{align}
\paragraph{}
What remains is to determine the asymptotic limits of the
$\eta$-linearised projectors $\mathcal{P}_{N+1}[\tilde w]$ and
$\mathcal{Q}_{N+1}[\tilde w]$. These involve
$\Psi_N(r)|_{\pm\infty}$ and $\Psi_N(1/r)|_{\pm\infty}$
respectively, which can be expressed in terms of the asymptotic
limits of $\Psi_0(r) $ and $\Psi_0(1/r)$ by applying the dressing
method iteratively, using the simplified expressions
(\ref{projectores1},\ref{projectores2}) for the lower
order projectors that we have found out. 
Explicitly we have
\begin{align}\label{plim1}
	\mathcal{P}_{N+1}[\tilde w]
	\Big|_{\pm\infty}&=\Psi_N(r)\Big|_{\pm\infty} W_\mathcal{P}[\tilde w]\, 
	\bar\Psi_N(r)\Big|_{\pm\infty},
	\\\label{plim2}
	\mathcal{Q}_{N+1}[\tilde w]
	\Big|_{\pm\infty}&=\Psi_N(1/r)\Big|_{\pm\infty} W_\mathcal{Q}[\tilde
	w]\bar\Psi_N(1/r)\Big|_{\pm\infty}.
\end{align}
where
\[ W_\mathcal{P}[\tilde w]=\frac{\tilde{w}\otimes
\tilde{w}^\dagger}{\tilde{w}\cdot
\tilde{w}^\dagger}\quad\text{and}\quad
W_\mathcal{Q}[\tilde w]=J\frac{\bar{\tilde{w}}\otimes
\tilde{w}^T}{\bar{\tilde{w}}\cdot \tilde{w}^T}J^{-1}.
\]
From \eqref{deltag}, (\ref{psinlim1}-\ref{psinlim2}) and
(\ref{plim1}-\ref{plim2}) one can easily determine the phase shifts
for a given polarisation $\tilde w$. The result is that the phase
shifts will always be additive, as expected from the factorisable
of the system: \textit{The total phase shift experienced by a
plane-wave scattering off a $N$-soliton background is equal to the sum
of the individual phase shifts caused by the scattering between a
plane wave and each constituent soliton.} 
\paragraph{}
For the two polarisation
types that we are considering, we have
\begin{eqnarray}
	W_\mathcal{P}^\parallel&\equiv& W_\mathcal{P}[\tilde
	w_\parallel]=\frac12\begin{pmatrix}
	1 & i & 0 & 0 \\
	-i & 1 & 0 & 0 \\
	0 & 0 & 0 & 0 \\
	0 & 0 & 0 & 0
	\end{pmatrix},\quad
	W_\mathcal{Q}^\parallel\equiv W_\mathcal{Q}[\tilde
	w_\parallel]=\frac12\begin{pmatrix}
	0 & 0 & 0 & 0 \\
	0 & 0 & 0 & 0 \\
	0 & 0 & 1 & i \\
	0 & 0 & -i & 1
	\end{pmatrix},\nn\\
	W_\mathcal{P}^\perp&\equiv& W_\mathcal{P}[\tilde
	w_\perp]=\frac12\begin{pmatrix}
	1 & 0 & 0 & i \\
	0 & 0 & 0 & 0 \\
	0 & 0 & 0 & 0 \\
	-i & 0 & 0 & 1
	\end{pmatrix},\quad
	W_\mathcal{Q}^\perp\equiv W_\mathcal{Q}[\tilde
	w_\perp]=\frac12\begin{pmatrix}
	0 & 0 & 0 & 0 \\
	0 & 1 & i & 0 \\
	0 & -i & 1 & 0 \\
	0 & 0 & 0 & 0
	\end{pmatrix}.
	\end{eqnarray}
This will give for $\tilde w=\tilde w_\parallel$,
\begin{align}
	\delta Z_1\Big|_{+\infty} &=
	-i\sin\Bigl(\frac{p}2\Bigr)e^{iv/2},\qquad \delta Z_1\Big|_{-\infty}
	= -i\sin\Bigl(\frac{p}2\Bigr)e^{-iP}e^{iv/2},\\
	\delta Z_2\Big|_{+\infty} &=
	\sin\Bigl(\frac{p}2\Bigr)e^{-iP}e^{iv/2}\prod_{j=1}^N\frac{r-X_j^+}{r-X_j^-},
	\qquad \delta Z_2\Big|_{-\infty} =\sin\Bigl(\frac{p}2\Bigr)e^{iv/2}
	\prod_{j=1}^N\frac{r-X_j^-}{r-X_j^+},
	\\
	\delta Z_3\Big|_{+\infty} &=\delta Z_3\Big|_{-\infty}=0,
\end{align}
where $P=\sum_{j=1}^N p_j$ is the total dyonic giant magnons momentum.
For $\tilde w=\tilde w_\perp$ we get,
\begin{align}
	\delta Z_1\Big|_{+\infty} &=
	-i\sin\Bigl(\frac{p}2\Bigr)e^{iv/2},\qquad \delta Z_1\Big|_{-\infty}
	= -i\sin\Bigl(\frac{p}2\Bigr)e^{-iP}e^{iv/2},\\
	\delta Z_2\Big|_{+\infty} &=\delta Z_2\Big|_{-\infty}=0,\\
	\delta Z_3\Big|_{+\infty}
	&=\sin\Bigl(\frac{p}2\Bigr)e^{iv}\prod_{j=1}^N\frac{1/r-X^-_j}{1/r-X^+_j},\qquad
	\delta Z_3\Big|_{-\infty}
	=\sin\Bigl(\frac{p}2\Bigr)e^{iv}\prod_{j=1}^N\frac{r-X^+_j}{r-X^-_j},
\end{align}
\paragraph{}
Here we list the resultant phase shifts constructed from dressing
method for the scattering between a plane wave and a general
$N$-soliton configuration lying within a given $S^{3}\subset S^{5}$,
parameterised by $|Z_{1}|^{2}+|Z_{2}|^{2}=1$. For the plane wave
perturbations that are parallel to the $S^3$ subspace, with
polarisation vector $\tilde w_\parallel=(i,1,0,0)^T$, we obtain
\begin{align}
	\delta Z_1,\delta\bar Z_1&:\quad    \delta_1\left(r;\{X_j^{\pm}\}\right)
	=-\delta_{\bar 1}\left(1/r,\{X_{j}^{\pm}\}\right)=P,\label{PaZ1}\\
	\delta Z_2,\delta\bar Z_2&:\quad \delta_2\left(r;\{
	X_j^\pm\}\right)=-\delta_{\bar2}\left(1/r,\{X_j^{\pm}\}\right)=
	-2i\sum_{j=1}^N\log\left(\frac{r-
	X_j^+}{r- X_j^-}\right)-P,\label{PaZ2}\\
	\delta Z_3,\delta\bar Z_3&:\quad\delta_3\left(r;\{X_j^{\pm}\}\right)=
	-\delta_{\bar 3}\left(1/r,\{X_j^{\pm}\}\right)=0.\label{PaZ3}
	\end{align}
If we take the giant magnon limit on the $N$-solitons
$X_j^{\pm}\rightarrow x^\pm_j\equiv e^{\pm ip_j/2}$
the expressions above reduce to,
\begin{align}
	\delta Z_1,\delta \bar Z_1&:\quad\delta_1\left(r;\{x_j\}\right)
	=-\delta_{\bar 1}\left(1/r,\{x^\pm_j\}\right)=P,\label{GMPaZ1}\\
	\delta Z_2,\delta \bar Z_2&:\quad\delta_2(r;\{x^\pm_j\})
	=-\delta_{\bar 2}(1/r,\{x^\pm_j\})=-2i\sum_{j=1}^N\log\left(\frac{
	r-\xp_j}{r-\xm}\right)-P,\label{GMPaZ2}\\
	\delta Z_3,\delta \bar Z_3&:\quad\delta_3\left(r;\{x^\pm_j\}\right)
	=-\delta_{\bar 3}\left(1/r,\{x^\pm_j\}\right)=0.\label{GMPaZ3}
	\end{align}
For the perturbations that are transverse to the $S^{3}$
but within $S^{5}$, with the polarisation vector $\tilde
w_\perp=(i,0,0,1)^T$, we obtain
\begin{align}
    \delta Z_1,\delta\bar Z_1&:\quad\delta_1\left(r;\{X_j^{\pm}\}\right)
	=-\delta_{\bar 1}\left(1/r,\{X_j^{\pm}\}\right)=P,\label{TraZ1}\\
    \delta Z_2,\delta \bar Z_2&:\quad\delta_2\left(r;\{X_j^{\pm}\}\right)
	=-\delta_{\bar 2}\left(1/r,\{X_j^{\pm}\}\right)=0,\label{TraZ2}\\
    \delta Z_3,\delta \bar Z_3&:\quad\delta_3\left(r;\{X_j^{\pm}\}\right)=
	-\delta_{\bar 3}\left(1/r,\{X_j^{\pm}\}\right)
	=-i\sum_{j=1}^N\log\left(\frac{r-
	X_j^+}{r-X_j^-}\right)
	 -i\sum_{j=1}^{N}\log\left(\frac{1/r- X_j^-}{1/r- X_j^+}\right).\label{TraZ3}
\end{align}
Notice that $\delta_3(r;\{X_j^\pm\})=\delta_{\bar 3}(r;\{X_j^\pm\})$
for this polarisation.
In the giant magnon limit these expressions again reduce to,
\begin{align}
	\delta Z_1,\delta \bar Z_1 &:\quad \delta_1\left(r;\{x^\pm_j\}\right)
	=-\delta_{\bar{1}}\left(1/r;\{x^\pm_j\}\right)=P,\label{GMTraZ1}\\
	\delta Z_2,\delta\bar Z_2 &:\quad \delta_2\left(r;\{x^\pm_j\}\right)
	=-\delta_{\bar 2}\left(1/r;\{x^\pm_j\}\right)=0,\label{GMTraZ2}\\
	\delta Z_3,\delta\bar Z_3 &:\quad \delta_3\left(r;\{x^\pm_j\}\right)
	=-\delta_{\bar
	3}\left(1/r,\{x^\pm_j\}\right)=-2i\sum_{j=1}^N\log\left(
	\frac{r-\xp}{r-\xm}\right)-P.\label{GMTraZ3}
\end{align}
\paragraph{}
For the main calculation of this paper, only the non-trivial
phase-shifts will be important. We present them again in a more
convenient notation. A perturbation with $\tilde w\equiv w_\parallel$
will correspond, as it was said, to a plane-wave travelling in a
direction parallel to the direction where the background solitons are
moving, i.e, in $Z_2$ and $\bar Z_2$. Only in these directions we will
have a non-trivial phase shift from the scattering for this particular
polarisation $\tilde w=w_\parallel$.  Hence we will label these by
$\delta_{Z_2}$ and $\delta_{\bar Z_2}$ to refer to a plane-wave
travelling along these directions. In the same fashion, for a
perturbation with $\tilde w\equiv w_\perp$ the scattering will occur
in the perpendicular directions $Z_3$ and $\bar Z_3$ to the moving
solitons, and so the phase shifts will be represented by
$\delta_{Z_3},\delta_{\bar Z_3}$. 
\begin{align}
	\delta_{Z_2}\left(r;\{
	X_j^\pm\}\right)&=-\delta_{\bar Z_2}\left(1/r,\{X_j^{\pm}\}\right)
	=-2i\sum_{j=1}^N\log\left(\frac{r-
	X_j^+}{r- X_j^-}\right)-P,\\
	\delta_{Z_3}\left(r;\{X_j^{\pm}\}\right)&=
	\delta_{\bar Z_3}\left(r;\{X_j^{\pm}\}\right)
	=-i\sum_{j=1}^N\log\left(\frac{r-
	X_j^+}{r-X_j^-}\right)
	 -i\sum_{j=1}^{N}\log\left(\frac{1/r- X_j^-}{1/r- X_j^+}\right).
\end{align}
In the GM limit $X_j^{\pm}\approx \exp(\pm i p_j/2)$ these take the
form,
\begin{align}
	\delta_{Z_2}(r;\{x^\pm_j\})&=-\delta_{\bar Z_2}(1/r;\{x^\pm_j\})=-2i\sum_{j=1}^N\log\left(\frac{
	r-\xp}{r-\xm}\right)-P,\\
	\delta_{Z_3}\left(r;\{x^\pm_j\}\right)&=\delta_{\bar
Z_3}\left(r;\{x^\pm_j\}\right)=
-2i\sum_{j=1}^N\log\left(\frac{r-\xp}{r-\xm}\right)-P.
\end{align}

\subsection{The explicit $\fsu(2|2)$ S-matrix}\label{ExpSmatrix}
\paragraph{}
Here we write out the explicit form for the $\fsu(2|2)$ dynamic
S-matrix entering in (\ref{fullSmatrix}), following the notations used
in \cite{AFZ1} (See also \cite{KMRZ})
\begin{eqnarray}
\his(x,y)&=&a(x,y)(E_1^1\otimes E_1^1\prime+E_2^2\otimes
E_2^2\prime+E_1^1\otimes E_2^2\prime+E_2^2\otimes E_1^1\prime)\nn\\
&+&b(x,y)(E_1^1\otimes E_2^2\prime+E_2^2\otimes E_1^1\prime-
E_1^2\otimes E_2^1\prime-E_2^1\otimes E_2^1\prime)\nn\\
&+&c(x,y)(E_3^3\otimes E_3^3\prime+E_4^4\otimes E_4^4\prime+
E_3^3\otimes E_4^4\prime+E_4^4\otimes E_3^3\prime)\nn\\
&+&d(x,y)(E_3^3\otimes E_4^4\prime+E_4^4\otimes E_3^3\prime-
E_4^3\otimes E_3^4\prime-E_3^4\otimes E_4^3\prime)\nn\\
&+&e(x,y)(E_1^1\otimes E_3^3\prime+E_1^1\otimes E_4^4\prime+
E_2^2\otimes E_4^4\prime+E_2^2\otimes E_4^4\prime)\nn\\
&+&f(x,y)(E_3^3\otimes E_1^1\prime+E_4^4\otimes E_1^1\prime+
E_3^3\otimes E_2^2\prime+E_4^4\otimes E_2^2\prime)\nn\\
&+&g(x,y)(E_1^4\otimes E_2^3\prime+E_2^3\otimes E_1^4\prime-
E_2^4\otimes E_1^3\prime-E_1^3\otimes E_2^4\prime)\nn\\
&+&h(x,y)(E_3^2\otimes E_4^1\prime+E_4^1\otimes E_3^2\prime-
E_4^2\otimes E_3^1\prime-E_3^1\otimes E_4^2\prime)\nn\\
&+&k(x,y)(E_3^1\otimes E_1^3\prime+E_4^1\otimes E_1^4\prime+
E_2^3\otimes E_2^3\prime+E_4^2\otimes E_2^4\prime)\nn\\
&+&l(x,y)(E_3^1\otimes E_1^3\prime+E_4^1\otimes E_1^4\prime+
E_3^2\otimes E_2^3\prime+E_4^2\otimes E_2^4\prime)\,.
\label{ExplicitSmatrix}
\end{eqnarray}
The various components in (\ref{ExplicitSmatrix}) for two magnons with
spectral parameters $x^{\pm}$ and $y^{\pm}$ are given by
\begin{eqnarray}
a(x,y)&=&\frac{x^+-y^-}{x^--y^+}\frac{\eta_y\eta_x}{\tilde\eta_y\tilde\eta_x}\,,~~
b(x,y)=\frac{(y^--y^+)(x^--x^+)(y^++x^+)}{(x^+-y^-)(y^-x^--y^+x^+)}
\frac{\eta_x\eta_y}{\tilde\eta_x\tilde\eta_y}\,,\nn\\
c(x,y)&=&-1\,,~~~
d(x,y)=\frac{(y^--y^+)(x^--x^+)(y^++x^+)}{(y^--x^+)(y^-x^--y^+x^+)}\,,\nn\\
e(x,y)&=&\frac{y^--x^-}{y^+-x^-}\frac{\eta_x}{\tilde\eta_x}\,,~~~
f(x,y)=\frac{x^+-y^+}{x^--y^+}\frac{\eta_y}{\tilde\eta_y}\,,\nn\\
g(x,y)&=&i\frac{(y^--y^+)(x^--x^+)(x^+-y^+)}{(x^+-y^-)(y^-x^--y^+x^+)
\tilde\eta_y\tilde\eta_x}\,,~~~
h(x,y)=i\frac{\ym\xm}{\yp\xp}\frac{(\ym-\yp)(\xm-\xp)(\xp-\yp)}{\eta_y\eta_x(\ym-\xp)(1-\ym\xm)}\,,\nn\\
k(x,y)&=&\frac{\xp-\xm}{\xm-\yp}\frac{\eta_y}{\tilde{\eta}_x}\,,~~~l(x,y)
=\frac{\yp-\ym}{\xm-\yp}\frac{\eta_{x}}{\tilde{\eta}_y}\,.
\label{ExplicitComponents}
\end{eqnarray}
The functions $\eta_{x}, \eta_{y}, \tilde{\eta}_{x}$ and
$\tilde{\eta}_{y}$ are used to account for the difference between the
``gauge/spin-chain'' and the ``string'' basis:
\begin{eqnarray}
{\rm Gauge}&\colon& \eta_{x}=\tilde{\eta}_{x}=\sqrt{i(\xm-\xp)}\,,~~~\eta_{y}
=\tilde{\eta}_{y}=\sqrt{i(\ym-\yp)}\,;\label{GBasis}\\
{\rm String}&\colon& \eta_{x}=\tilde{\eta}_{x}\sqrt{\frac{\yp}{\ym}}
=\sqrt{i(\xm-\xp)\frac{\yp}{\ym}}\,,~~~
\eta_{y}=\tilde{\eta}_{y}\sqrt{\frac{\xm}{\xp}}=
\sqrt{i(\ym-\yp)\frac{\xm}{\xp}}\,.\label{SBasis}
\end{eqnarray}
Essentially, if we choose the gauge basis (\ref{GBasis}), the
components in (\ref{ExplicitComponents}) are the same as the ones in
\cite{BeisertSU22}). However as we would like to compare the
semiclassical phase shifts with the results obtained from the string
sigma model calculations, it is in fact necessary for us to select the
string basis (\ref{SBasis}) to obtain the exact matches.

\section{Useful Integrals for the Evaluation of Semiclassical
Phase}\label{Integrals}

Here we list the useful integrals for evaluating the semiclassical
phase, using the formula (\ref{oneloopformula}):
\begin{eqnarray}
&&\int^{+1}_{-1}\,dr\frac{1}{r-a}\frac{1}{r-b}
=\frac{1}{a-b}\left[\log\left(\frac{a-1}{a+1}\right)-
\log\left(\frac{b-1}{b+1}\right)\right]\label{Aint1}\,,\\
&&\int^{+}_{-1}\,dr \frac{1}{r-a}\frac{1}{b-1/r}
=\frac{1}{b-1/a}\log\left(\frac{a-1}{a+1}\right)-
\frac{1}{b(ab-1)}\log\left(\frac{1-b}{1+b}\right)\label{Aint2}\,.
\end{eqnarray}

%%%%%%%%%%%%%%%%%%%%%%%%%
%%%%%%%%%%%%%%%%%%%%%%%%%
\bibliographystyle{unsrt}

\end{document}